\documentclass[a4paper]{article}
\usepackage{listings}
\usepackage{xcolor}
\usepackage{multirow}
\usepackage{jheppub}  
\usepackage{natbib}
\usepackage{dcolumn}
\usepackage[english]{babel}
\usepackage[utf8x]{inputenc}
\usepackage[T1]{fontenc}
\usepackage{palatino}

\usepackage{amsmath}
\usepackage{afterpage}
\usepackage{graphicx, subcaption}
\usepackage[colorinlistoftodos]{todonotes}
\usepackage[colorlinks=true]{hyperref}
\usepackage{makeidx}
\newcommand{\be}{\begin{equation}}  
\newcommand{\ee}{\end{equation}}  
\newcommand{\bea}{\begin{eqnarray}}  
\newcommand{\eea}{\end{eqnarray}}
  
\makeindex
\begin{document}

\vspace*{1.2cm}

\thispagestyle{empty}
\begin{center}

\par\vspace*{7mm}\par

{\LARGE \bf Central exclusive production of $W$ boson pairs in $pp$ collisions at the LHC in hadronic and semi-leptonic final states}

\par\vspace*{20mm}\par

{\large \bf C. Baldenegro$^a$, G. Biagi$^{a,b}$, G. Legras$^{a,c}$, C. Royon$^a$}

\bigskip

{\em $^a$ The University of Kansas, Lawrence, Kansas, U.S.}
\\
{\em $^b$ École Centrale, Paris, France}
\\
{\em $^c$ École des Mines-Paristech, Paris, France}
\vspace*{5mm}

{\em E-mail: 
c.baldenegro@cern.ch\\
biagiguillaume@gmail.com\\
gauthier.legras@mines-paristech.fr\\
christophe.royon@cern.ch}

\vspace*{15mm}

{  \bf  Abstract }

\end{center}

\vspace*{1mm}

\begin{abstract}

We present a phenomenology study on central exclusive production of $W^+W^-$ boson pairs in proton-proton collisions at the Large Hadron Collider at 14 TeV using the forward proton detectors, such as the ATLAS Forward Proton or the CMS-TOTEM Precision Proton Spectrometer detectors. Final states where at least one of the $W$ bosons decay hadronically in a large-radius jet are considered. The latter extends previous efforts that consider solely leptonic final states.
A measurement of exclusive $W^+W^-$ also allows us to further constrain anomalous quartic gauge boson interactions between photons and $W$ bosons. Expected limits on anomalous quartic gauge couplings $a_{0,C}^W$ associated to dimension-six effective operators are derived for the hadronic, semi-leptonic, and leptonic final states. It is found that the couplings can be probed down to one-dimensional values of $a_{0}^W = 3.7\times 10^{-7}$ GeV$^{-2}$ and $a_{C}^W = 9.2 \times 10^{-7}$ GeV$^{-2}$ at $95\%$ CL at an integrated luminosity of 300 fb$^{-1}$ by combining all final states, compared to values of about $a_{0}^W = 4\times 10^{-6}$ GeV$^{-2}$ and $a_{C}^W = 1\times 10^{-5}$ GeV$^{-2}$ at 95\% CL expected for the leptonic channel alone.

\end{abstract}

\clearpage

\section{Introduction}

One of the main goals of the CERN Large Hadron Collider (LHC) program is the discovery of physics beyond the Standard Model (SM) of particle physics. In various extensions of the SM, additional interactions or particles at higher energies are expected to have an impact on the electroweak sector of the SM~\cite{AQGC_FichetGersdorff,EFT1}. Particular examples of SM extensions that induce can induce deviations from SM predictions of $\gamma\gamma\rightarrow W^+W^-$ scattering include composite Higgs models or warped extra-dimensions, for instance~\cite{AQGC_FichetGersdorff,Espriu:2014jya,Delgado:2014jda}. It is thus important to understand the couplings between the gauge bosons of the weak and electromagnetic interactions of the SM.

\begin{figure}
\begin{center}
\begin{subfigure}{0.45\textwidth}
\includegraphics[scale=.5]{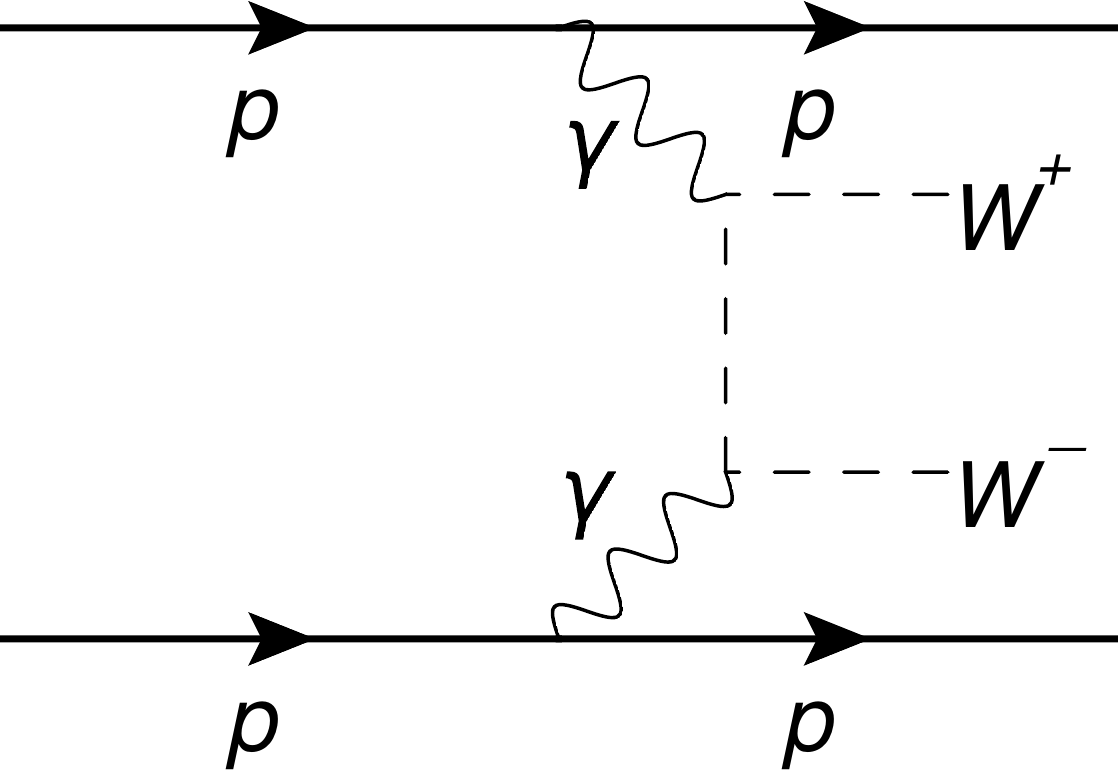}
\end{subfigure}
\begin{subfigure}{0.45\textwidth}
\includegraphics[scale=.5]{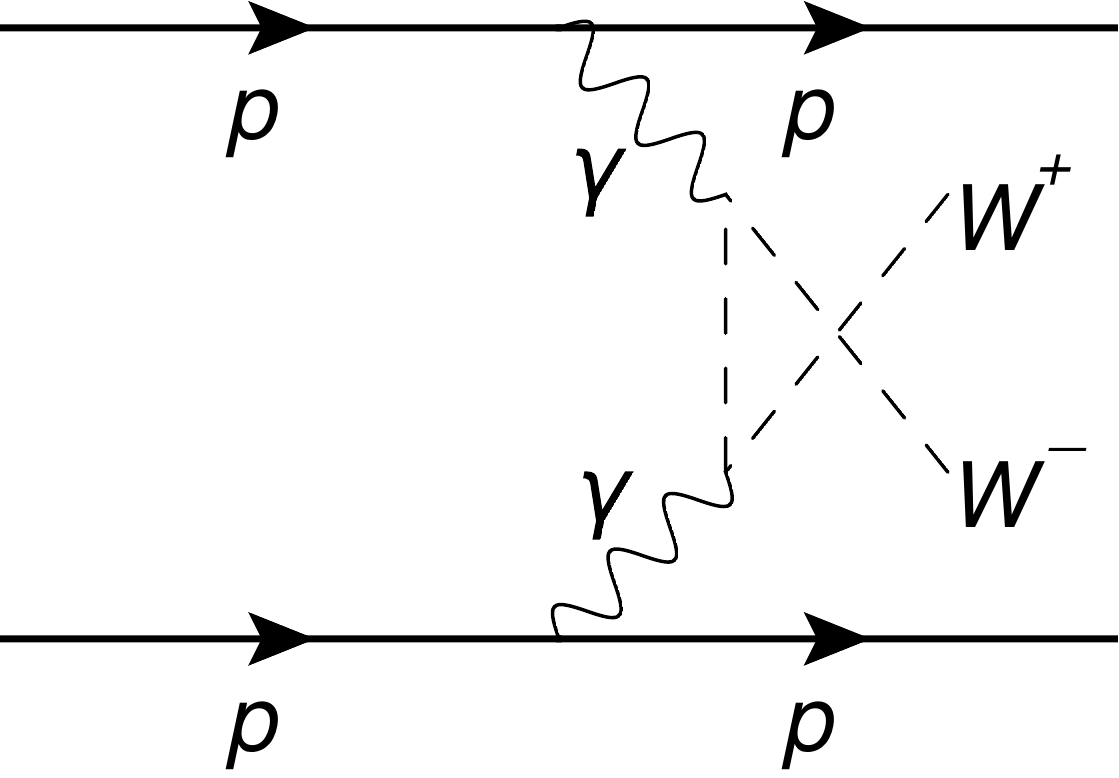}
\end{subfigure}
\begin{subfigure}{0.45\textwidth}
\includegraphics[scale=.5]{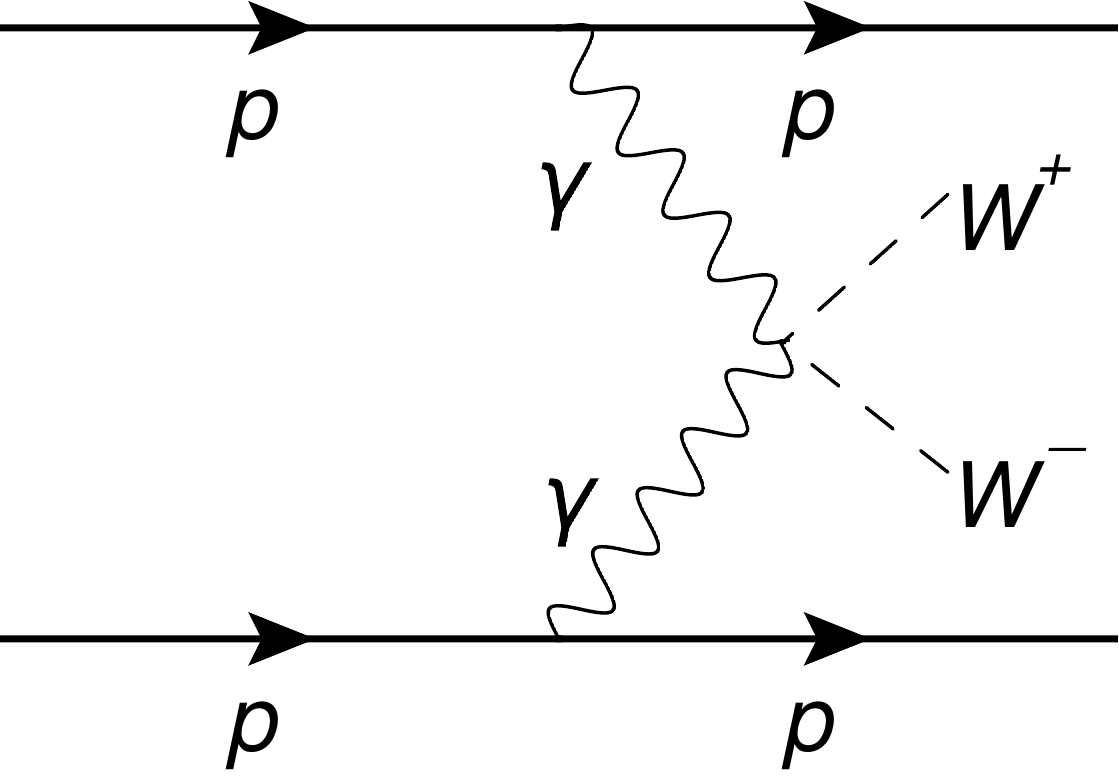}
\end{subfigure}
\caption{\label{fig:diagrams} Leading order SM diagrams contributing to $pp\rightarrow p W^+W^- p$. The upper two diagrams represent the $t$- and $u$-channels contribution induced by the SM $\gamma W^+W^-$ coupling, whereas the lower diagram corresponds to the SM $\gamma\gamma W^+W^-$ coupling in the SM.}
\end{center}
\end{figure}

By virtue of the non-abelian nature of electroweak sector of the SM, direct interaction couplings between the vector gauge bosons are possible. The LHC experiments have gathered evidence for some of these non-abelian couplings, while others are very difficult to probe. Recent reviews on these topics can be found in Refs.~\cite{VBSreview, Gallinaro:2020cte}. In this paper, we are mostly interested in the interactions between photons and $W$ bosons, which interact via triple ($\gamma W^+ W^-$) and quartic ($\gamma\gamma W^+ W^-$) couplings in the SM. These couplings are present in the SM at tree-level, and are fully connected through the requirement of gauge invariance in the SM. These interactions can be directly probed via two-photon fusion $\gamma\gamma\rightarrow W^+W^-$ in pp collisions, as shown in Fig.~\ref{fig:diagrams}. This process is the main topic of discussion in this paper.

In high energy proton-proton (pp) collisions, the electromagnetic fields generated by the relativistic protons can be treated as a source of quasi-real photons~\cite{WeiszackerWilliams,Budnev}. Thus, in addition to the standard quark or gluon exchanges in pp collisions, one can study reactions with photon exchange off the proton at the LHC, and use this to study photon-photon collisions at high energies. In some of these interactions with quasi-real photon exchange, the proton may survive the interaction, and will be scattered afterwards at very small angles with respect to the beam. The scattered protons can be tagged with near-beam tracking detectors, known as Roman Pots (RPs), located at about 200 m with respect to the interaction point. The ATLAS and CMS-TOTEM Collaborations have added RP detectors during Run-2 at the LHC, known as ATLAS Forward Proton (AFP) and CMS-TOTEM Precision Proton Spectrometer (PPS) ~\cite{AFP,PPS}. Thus, the $\gamma\gamma\rightarrow W^+W^-$ process can be studied in central exclusive production of $W$ boson pairs $pp\rightarrow p W^+W^- p$~\cite{ggWW_ChaponKepkaRoyon,PhysRevD.78.073005}. Previous phenomenology studies based on the use of the proton tagging technique to study two-photon fusion interactions can be found in Refs.~\cite{yyyZ_Baldenegro, yyyy_Fichet, Fichet:2016clq, Inan:2014mua, Senol:2014vta, Fichet:2016pvq, Gupta:2011be, Senol:2013ym,Sun:2014qoa,Fichet:2013gsa,Sun:2014qba,TAHERIMONFARED2016301,Gurkanli:2020tee,PhysRevD.78.073005,Sylvain_PRD,Baldenegro_ALP, Goncalves:2020saa, Harland-Lang:2020veo}. Other interesting studies that aim also to describe photon fluxes in processes with proton dissociation, treated in the $k_T$ factorization framework, are presented in Refs.~\cite{Szczurek:2019ihz, Luszczak:2018ntp, Forthomme:2018sxa}. Since our studies rely on events with proton tagging, no such treatment is necessary. Experimentally, the ATLAS and CMS-TOTEM Collaborations have observed $\gamma\gamma\rightarrow \ell^+\ell^-$ using the proton tagging technique~\cite{dilepton_PPS,dilepton_AFP}, and the CMS-TOTEM Collaborations reported a first search of central exclusive $\gamma\gamma$ production using the proton tagging technique~\cite{exclusive_diphoton_PPS}.

Early searches sensitive to $\gamma\gamma W^+W^-$ couplings were done at the CERN LEP electron-positron collider by the OPAL and DELPHI Collaborations in $W^+W^-\rightarrow \gamma\gamma$ scattering ~\cite{WWyy_LEP,WWyy_LEP2,WWyy_LEP3,WWyy_LEP4}, followed by the D0 experiment at the Fermilab Tevatron in proton-antiproton collisions at $\sqrt{s} = 1.96$ TeV\cite{Abazov:2013opa} and by the CMS and ATLAS experiments at the CERN LHC~\cite{CMS_yyWW7TeV,CMS_yyWW8TeV, ATLAS_yyWW8TeV} at $7$ and $8$ TeV in central exclusive production of $W^+W^-$ pairs in pp collisions ($pp\rightarrow p^{(*)} (\gamma\gamma\rightarrow W^+W^-) p^{(*)}$), the protons may dissociate into a low mass excited state or remain intact. In a recent conference note~\cite{gammagammaWW_ATLASCONF}, the ATLAS Collaboration reported the observation of $\gamma\gamma\rightarrow W^+W^-$ scattering at 8.4$\sigma$ in the $e^\pm \mu^\mp$ channel without proton tagging, consistent with the SM expectations. In order to obtain a robust understanding of this scattering process, one needs to consider an expansion in the search strategy, especially for a better understanding of in the high invariant mass regime. As an additional advantage, one can use these as standard candle processes for the proton spectrometer calibration, acceptance and efficiency determination, of crucial importance for physics searches based on proton tagging.

In previous experimental and phenomenological studies, only leptonic decays of each $W$ boson have been considered in central exclusive production. Indeed, purely leptonic final-states leave a very clean signature in the detectors: two leptons ($e^+e^-$, $\mu^+\mu^-$, $e^\pm\mu^\mp$) with an amount of missing transverse momentum $p_T^\text{miss}$ associated to the transverse momentum carried away by the undetected neutrinos. Such a signature has allowed us to study central exclusive $W^+W^-$ with a moderate number of simultaneous pp collisions per bunch crossing (pileup). However, the visible cross section is largely reduced due to the small branching fraction of the $W$ boson decay into leptons (less than 5\% of $W$ boson pairs decay into muon or electron flavored leptons). Thus, in the interest of better understanding this electroweak gauge boson scattering process within the SM in a wider kinematic range, and also to enhance our chances of discovering physics beyond the SM, one can consider the decay of the $W$ boson into quark-antiquark pairs, which has a much larger branching fraction. The study with at least one $W\rightarrow q\bar{q}'$ decay has a larger background contribution from standard mechanisms of jet production in quantum chromodynamics, which need to be quantified. New strategies need to be developed in order to use these channels.

The production cross section can increase significantly if we consider hadronic decays of at least one of the $W$ bosons, allowing for better sensitivity for the central exclusive $W^+W^-$ boson decay to be identified experimentally. If the $W$ boson is highly boosted, the hadronic decay of a single $W$ boson can result in a single large-radius jet whose mass is compatible with that of the $W$ boson. In this paper, we consider fully-hadronic (JJ), semi-leptonic (J+$\ell\nu$), and fully-leptonic ($\ell\nu\bar{\ell}\bar{\nu}$) decays of the $W^+W^-$ system in central exclusive production, where $J$ is a large-radius jet, $\ell$ is either a muon or an electron, and $\nu$ the neutrino. The branching ratio of $W^+W^-$ pair decay in hadronic, semi-leptonic, and leptonic channels is, in principle, about 46\%, 29\%, and 4.7\%, respectively (considering electrons or muons for leptonic decays where the remaining 20\% of the branching fraction corresponds to decays in tau leptons). In the present paper, as an extension of the study in Refs.~\cite{ggWW_ChaponKepkaRoyon,PhysRevD.78.073005}, we explicitly include the contribution of pileup backgrounds since our goal is to perform this measurement in standard pileup luminosity at the LHC at an integrated luminosity of 300 fb$^{-1}$.

In addition to the SM study of exclusive $W^+W^-$ production, we consider scenarios where the presence of new particles at much larger energies may induce anomalous contributions to the $\gamma\gamma\rightarrow W^+W^-$ scattering. In our study, we parametrize these new physics contributions by means of an effective field theory approach. Our focus in this part of the phenomenology study is on the improvement of the sensitivity to the anomalous interaction couplings by studying high-mass exclusive $W^+W^-$ production in hadronic and semi-leptonic final states, compared to standard benchmark sensitivities in the purely leptonic final state.

The paper is organized as follows. The theory framework, as well as Monte Carlo simulated events used in the study, are discussed in Sec.~\ref{sec:theory_framework}. Section \ref{sec:signal_bkg_treatment} gives a description of signal and backgrounds considered in the paper. The analysis strategy tailored to study the SM $\gamma\gamma\rightarrow W^+W^-$ is described in Sec.~\ref{sec:analysis_strategy}, with the respective results of the analysis in Sec.~\ref{sec:sm_results}. Similarly, we study how deviations from the SM could manifest using an effective field theory framework in Sec.~\ref{sec:anomalous_coupling}, with the respective results and projections on sensitivity to anomalous couplings described in Sec.~\ref{sec:anomalous_results}. The summary of the study is presented in Sec.~\ref{sec:summary}.

\section{The $pp \rightarrow p W^+W^- p $ process at the LHC}\label{sec:theory_framework}
\subsection{Equivalent photon approximation}\label{sec:epa}

We use the equivalent photon approximation~\cite{Budnev,WeiszackerWilliams} to describe the $pp \rightarrow p W^+W^- p $ process via coherent photon exchanges as discussed in~\cite{ggWW_ChaponKepkaRoyon,PhysRevD.78.073005,Terazawa}. In this approximation, the quasi-real photons are emitted by the incoming protons producing a state $X$ through photon fusion $\gamma\gamma\rightarrow X$, such as that shown in Fig.~\ref{fig:diagrams}. The quasi-real photon spectrum of virtuality $Q^2$ and energy $E_\gamma$ is given by:

\begin{equation}
dN = \frac{\alpha_{\rm em}}{\pi} \frac{\mathrm{d}E_\gamma}{E_{\gamma}} \frac{\mathrm{d}Q^2}{Q^2}\bigg[ \bigg(1 - \frac{E_\gamma}{E}\bigg) \bigg(1-\frac{Q^2_{\mathrm{min}}}{Q^2}\bigg )F_E + \frac{E_\gamma^2}{2E^2} F_M \bigg  ]
\end{equation}

where $E$ is the energy of the incoming proton of mass $m_p$, $Q^2_{\mathrm{min}}=m_p^2E_\gamma^2/[E(E-E_\gamma)]$ the photon minimum virtuality allowed by kinematics, $\alpha_\text{em}$ is the fine structure constant, and $F_E$ and $F_M$ are functions of the electric and magnetic form factors $G_E$ and $G_M$ of the proton. In the dipole approximation, the follow relationships hold

\begin{equation}
F_M=G^2_M  \qquad F_E=(4m_p^2G^2_E+Q^2G^2_M)/(4m_p^2+Q^2)\qquad G^2_E=G^2_M/\mu_p^2=(1+Q^2/Q^2_0)^{-4} \,.
\label{eq:newera:elmagform}
\end{equation}

The magnetic moment of the proton is $\mu_p^2=7.78$ and the fitted scale $Q^2_0=0.71$ GeV$^2$. Since the electromagnetic form factors fall steeply as a function of $Q^2$,
the cross section can be factorized into the matrix element of the photon fusion process and the two photon fluxes. In order to obtain the production cross section, the photon fluxes are first integrated over $Q^2$

\begin{equation}
f(E_\gamma)=\int^{\infty}_{Q^2_{\rm min}}\frac{d N}{d E_\gamma d Q^2} d Q^2 \,.
\label{sm:flux_q2}
\end{equation}
 The resulting photon energy spectra are given in Ref.~\cite{ggWW_ChaponKepkaRoyon,PhysRevD.78.073005}, for instance. These fluxes are used to define an effective two-photon luminosity 
$ d L^{\gamma\gamma} \slash d W$ obtained by integrating the two-photon fluxes
 \begin{eqnarray}
 f(E_{\gamma1})\,f(E_{\gamma2})\,d E_{\gamma1}\,dE_{\gamma2}\, \delta(W_{\gamma\gamma}-2\sqrt{E_{\gamma1} E_{\gamma2}})
 \end{eqnarray}

 \noindent where $W_{\gamma\gamma}$ is invariant mass of the initial-state photon pair. Using the effective photon luminosity, the integrated cross section for the $pp\rightarrow pXp $ process is given by
\begin{equation}
\sigma=\int\sigma_{\gamma\gamma\rightarrow X}
      \frac{\mathrm{d} L^{\gamma\gamma}}{\mathrm{d} W}d W 
\label{eq:sm:totcross}
\end{equation}
where $\sigma_{\gamma\gamma\rightarrow X}$ denotes the 
cross section of the sub-process $\gamma\gamma\rightarrow X$, dependent on the 
invariant mass of the two-photon system.
In addition to the photon exchange, there might be additional soft gluon exchanges that might destroy the protons. To take into account this effect, we can introduce the so-called survival probability that the protons remain intact in photon-induced processes~\cite{Khoze1}. The survival probability is of oder 1 in our case since we consider $WW$ exclusive production via QED processes, or photon exchanges~\footnote{It is worth noting that the survival probability is much smaller in the case of gluon exchanges where it is estimated to be about 0.03 at 14 TeV at the LHC.}.  In this paper, we assumed a uniform survival probability factor of order unity, namely 90\%~\cite{Khoze1,Khoze2,survival_khoze,survival_superchic2,survival_gotsman}, which is an approximation since there is a small dependence on the mass of the $WW$ system as illustrated for instance in Fig. 7 of Ref.~\cite{survival_khoze}.

\subsection{Forward proton detectors} \label{se:proton_det}

The $pp \rightarrow p W^+W^- p$ proccess can be probed with the detection of two intact protons in the forward proton detectors in the AFP or PPS~\cite{AFP,PPS}  and the reconstruction of the $W^+W^-$ boson decay products. The latter is discussed in Section~\ref{sec:wwdecay}. The forward detectors are located symmetrically at about 210 m from the main interaction vertex and cover a range of $0.015 < \xi_{1,2} < 0.15$ in CMS-TOTEM and ATLAS~\cite{AFP,PPS}, where $\xi \equiv \Delta p / p$ is the fractional momentum loss of the proton. This leads to an acceptance in the central diffractive mass $m_{X} = \sqrt{\xi_1 \xi_2 s}$ between $300$ and $2$ TeV, with $\sqrt{s} = 14$ TeV the center-of-mass energy used in our projections. The $\xi$ resolution is assumed to be of the order of 2\%, taking into account both beam related ans alignment uncertainties.

In central exclusive production processes, the forward scattered protons are kinematically correlated to the central system by virtue of four-momentum conservation. Indeed, the rapidity of the central system $y_{WW}$ and the rapidity reconstructed with the two tagged protons, $y_{pp} \equiv \frac{1}{2}\ln (\frac{\xi_1}{\xi_2})$, should be equal for central exclusive events, i.e., $y_{WW} = y_{pp}$. Likewise, the invariant mass reconstructed with the central detector is the same as that reconstructed with intact proton information, i.e., $m_{WW} = m_{pp}$ for central exclusive events. These relations can be used to get a strong background rejection, which is the key feature of the forward proton detectors in exclusive processes~\cite{yyyy_Fichet,yyyZ_Baldenegro}. Further background rejection can be achieved with the use of timing detectors. Timing detectors have been installed and operating in both  PPS and AFP to measure the time-of-flight of protons with a design precision of about 20 ps, which would allow to determine the interaction vertex of the protons with a 2 mm precision independently of the primary vertex reconstruction based on standard tracking techniques~\cite{PPS,AFP}. The time-of-flight information, with the aforementioned precision, allows for further pileup background rejection by a factor of up to $40$ by constraining the scattered proton to originate from the same vertex as the $W^+W^-$~\cite{PPS,AFP} for a pileup of 50 by comparing the vertex position from tracking to that determined by timing.

\subsection{Reconstruction of $W^+W^-$ decay products}\label{sec:wwdecay}

As mentioned in the introduction, we consider three decay channels of the $W^+W^-$ system: hadronic (two large-radius jets), semi-leptonic (one large-radius jet in association with an isolated lepton and $p_T^\text{miss}$), and purely leptonic ($\ell^+_1\ell^-_2$ and missing transverse momentum). Here, we consider that jets can be reconstructed within pseudorapidities of $|\eta|<5$, and consider electrons and muons to be reconstructed within pseudorapidities of $|\eta|<2.5$. For $W$ bosons decaying hadronically, the corresponding large-radius jet is clustered with the infrared and collinear safe anti-$k_t$ algorithm~\cite{Cacciari:2008gp} using particles at the stable particle level with the FastJet package~\cite{fastjet}, using a distance parameter of $R = \sqrt{\Delta\phi^2 + \Delta y^2} = 0.8$. The value of $R = 0.8$ is based on standard choices by the ATLAS and CMS Collaborations for large-radius jet analyses.

The phenomenology analysis is performed using stable particles at generator level. In order to mimic detector effects, we apply smearing effects on the transverse momenta and pseudorapidity and azimuthal angular of particles and jets reconstructed at generator-level. For jets, a conservative smearing of 2\% on jet transverse momentum is used based on Refs.~\cite{PFnoteCMS,PFnoteATLAS}. The latter is a combination of jet energy scale and jet energy resolution effects for large-radius jets. In addition, angular smearings of 1 mrad for azimuthal and polar angles are applied to each jet. The jet mass is computed at generator level and smeared by about 10\%, based on reports by the ATLAS and CMS Collaborations on jet mass resolution studies~\cite{WbosonBoostedATLAS,WbosonBoostedATLAS2,WbosonBoostedCMS}. The jet energy is also smeared by 10\%, and the longitudinal component of the jet momentum is determined following the energy-momentum relation.
For charged leptons in $|\eta| < 2.5$, a smearing of 2\% is applied on the reconstructed transverse momenta~\cite{leptonATLAS,leptonCMS} and additional angular smearings of 1 mrad for azimuthal and polar angles are applied to each lepton. The missing transverse momentum is defined at generator-level as the negative vector sum of the transverse momentum of each final-state particle within acceptance $|\eta|<5$. A smearing of 20\% is applied on the reconstructed missing transverse momentum, based conservatively on the observed performance of the ATLAS and CMS detectors in the reconstruction of this variable~\cite{MET_ATLAS,PFnoteCMS}.

In the last years, special attention has been given to the better understanding of large-radius jets and jet substructure techniques~\cite{jetSubstructureReview}. It has become increasingly important to distinguish large-radius jet objects that originate from the merging of the decay products of high-transverse momenta $W$ bosons ($W$ jet) from those initiated by light-flavor quarks or gluons in standard quantum chromodynamics interactions (QCD jets). The CMS and ATLAS Collaborations have presented results related to the decay of $W$ or $Z$ bosons using groomed and ungroomed large-radius jets in recent years~\cite{WbosonBoostedATLAS,WbosonBoostedCMS}. The most important kinematic discriminant is the invariant mass of the jet. The $W$ boson jet mass arises from the kinematics of the two jet cores associated to the decay and fragmentation of the two quarks, with some broadening induced by the soft-gluon emissions during the parton shower evolution and jet clustering effects. In contrast, the QCD jet mass arises mostly from soft-gluon radiation, resulting in a quickly falling distribution in the invariant mass of the jet. Thus, a selection requirement on the invariant mass of the large-radius jet around the mass of the $W$ boson already provides a robust selection requirement to suppress the contribution of QCD jets and to isolate $W$ boson large-radius jet candidates. Further discriminant variables based on jet substructure have been explored by the ATLAS and CMS experiments~\cite{jetSubstructureReview}. In this first study, we use a simple approach and choose not to apply these jet substructure techniques by default, and thus use the jet mass as the main jet substructure variable to identify the $W$ boson large-radius jet. This is because in central exclusive production, the events are very clean, not ``contaminated" by soft QCD radiation. Additional soft parton exchanges between proton remnants are absent (since the protons remain intact), as well as underlying event activity, and QCD initial-state radiation. These contributions need to be properly removed in standard $W$ boson large-radius jet analyses. The use of the discriminant variables introduced by ATLAS and CMS can only improve our performance studies, and need to be accounted for on top of the jet invariant mass for the experimental analysis.

\section{Signal and background treatment}\label{sec:signal_bkg_treatment}

\subsection{Standard model signal modelling}

Central exclusive production of $W^+W^-$ pairs is simulated with the Forward Physics Monte Carlo (FPMC) event generator~\cite{fpmc}. FPMC simulates diffractive exchanges and photon exchange in pp collisions. Photon exchange is simulated within the equivalent photon approximation described in Sec.~\ref{sec:epa}. The process $pp\rightarrow p W^+W^- p$ accounts for scattering amplitudes induced by the SM $\gamma W^+W^-$ and $\gamma\gamma W^+W^-$. These amplitudes were calculated with the CalcHEP package~\cite{calchep}, and coded into FPMC~\cite{Kepka:2008yx,PhysRevD.78.073005}. The SM cross section is 95.6 fb for central exclusive $W^+W^-$ production, and 5.9 fb for a $W^+W^-$ invariant mass above 1 TeV for $\sqrt{s} = 14$ TeV~\cite{Kepka:2008yx,PhysRevD.78.073005}.

\subsection{Background modelling}

The dominant backgrounds to exclusive $W^+W^-$ production originate from non-diffractive events that mimic the $W^+W^-$ decay signature in the central detector detected in coincidence with intact protons that originate from soft diffractive pileup events. Backgrounds related to single- and double-pomeron exchange $W^+W^-$ production lead to a negligible contribution in high-mass $W^+W^-$ pair production, as found in previous phenomenology analyses~\cite{ggWW_ChaponKepkaRoyon,PhysRevD.78.073005}, as well as in previous studies by ATLAS and CMS~\cite{CMS_yyWW7TeV, CMS_yyWW8TeV,ATLAS_yyWW8TeV, gammagammaWW_ATLASCONF}.

The average number of multiple pp collisions per bunch crossing sets a huge background environment on the search for central exclusive events at the LHC. Forward protons can be created in soft diffractive reactions, whose production cross section rates are expected to be of the order of $100$ mb at 14 TeV. During Run-2 of the CERN LHC, the typical number of pileup interactions ranged from 30 to 50 interactions per bunch crossing at the ATLAS and CMS interaction points during the standard luminosity fills. With this high amount of pileup interactions, and the large cross section of soft single- and central-diffractive processes in pp collisions, the amount of protons detected in the forward detectors is not negligible. As it will be shown later, this can be suppressed by exploiting the kinematic correlations between the forward two proton system and the central system. In all our studies presented in the following, we assume a conservative number of 50 pile up events per bunch crossing for our projections at $\sqrt{s} = 14$ TeV.

The dominant background corresponds to non-diffractive diboson production processes ($VV = W^+W^-$, $W^{\pm} Z$, $ZZ$), $W+$jets, $Z+$jets $t\bar{t}$, single-top, and QCD jets (jet production with quark and gluon strong interactions) detected in association with protons from uncorrelated pileup interactions in the same bunch crossing. In this paper, we refer to these backgrounds as $W^+W^-$+pileup, $ZZ$+pileup, $WZ$+pileup, $W$+jets+pileup, $Z$+jets+pileup, $t\bar{t}$+pileup, single-top+pileup, and QCD jets + pileup, respectively. The combination of all these backgrounds is referred to as ``pileup background'' throughout this paper. Standard QCD jets are the most important source of background in the fully-hadronic final state, as demonstrated later.

The diboson backgrounds are generated with the leading order HERWIG6 Monte Carlo event generator\cite{herwig6}, and the top quark and QCD jet backgrounds are simulated with the PYTHIA8 Monte Carlo event generator~\cite{pythia8}.

\begin{figure}
\centering
\includegraphics[width = 0.49\textwidth]{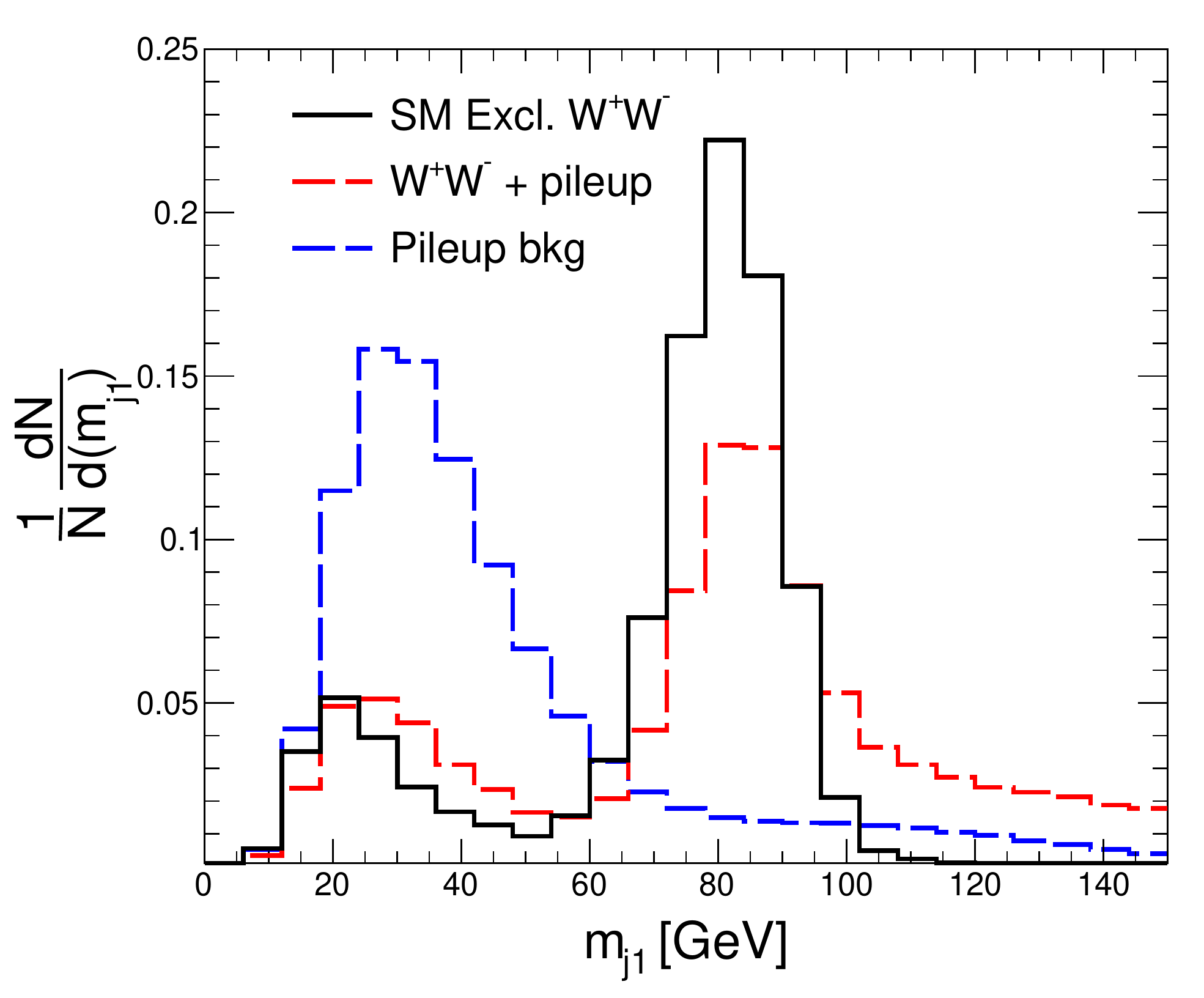}
\includegraphics[width=0.49\textwidth]{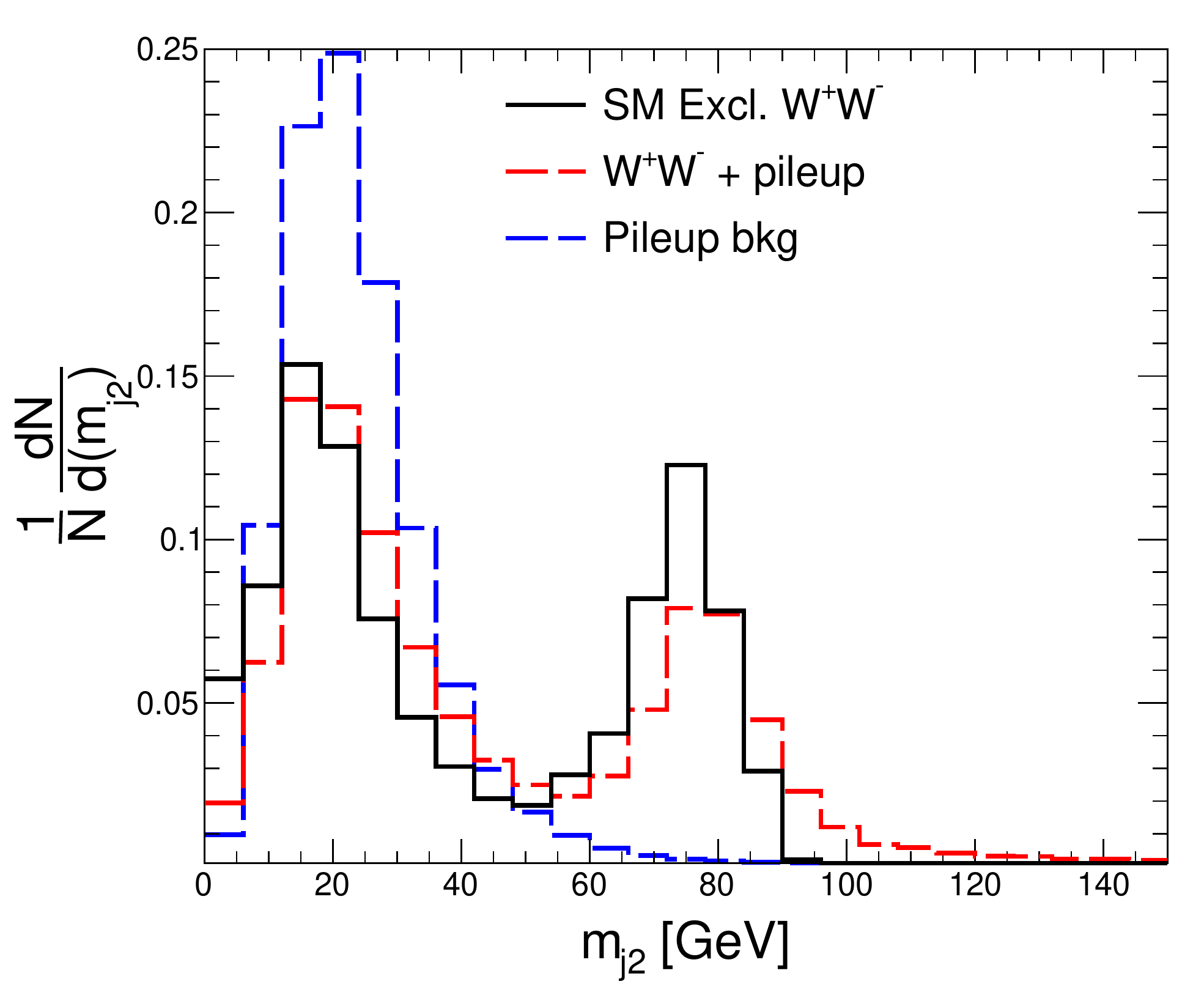}
\caption{Large-radius jet invariant mass distribution for the leading (left) and subleading (right) in mass in the hadronic channel. The distributions are normalized to unity. The SM exclusive $W^+W^-$ pair
production is displayed in black full line, the pileup background in blue dashed line  (after preselection requirements), and the pileup background subcomponent of $W^+W^-$+pileup is shown in red. The peak at around 80 GeV corresponds to the $W$ boson rest mass. The peak at lower masses corresponds to cases where only one of the quarks of the $W$ boson decay is tagged as a jet in the final state, or more generally for light-flavor quark jets or gluon jets.}
\label{SM-hadr-mass}
\end{figure}

\begin{figure}
\centering
\includegraphics[width=0.55\textwidth]{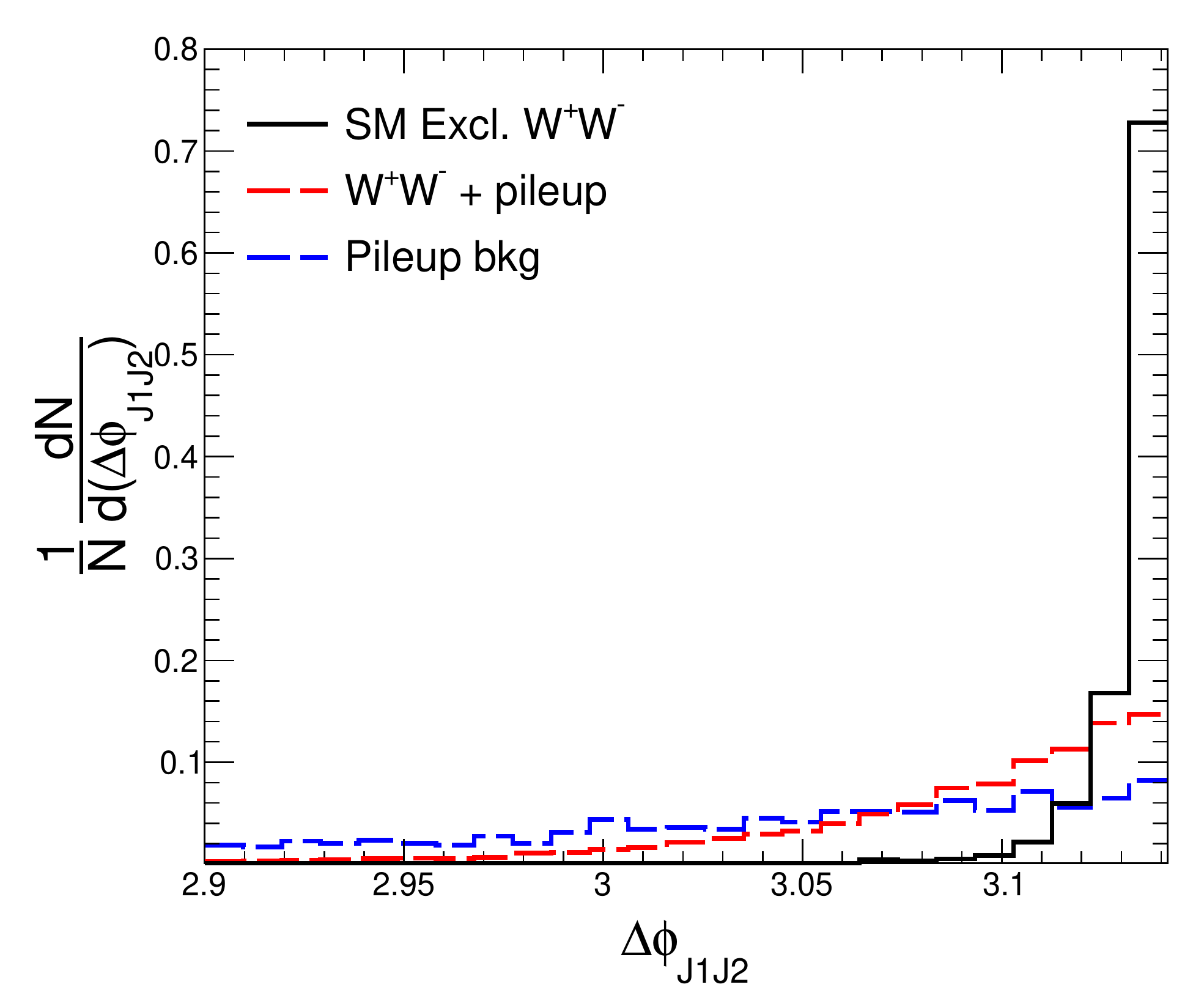}
\caption{Azimuthal angular difference between the two large-radius jets corresponding to the decay of the $W$ bosons, $\Delta \phi_\text{j1j2} = |\phi_\text{j1}-\phi_\text{j2}|$ after preselection and $m_\text{j1}$ and $m_\text{j2}$ requirements. SM central exclusive $W^+W^-$ events are represented by the black histogram, and non-diffractive $W^+W^-$+pileup background events are represented in red. The pileup background distribution (blue dashed histogram) is flatter due to the effect of initial-state radiation and underlying event activity in non-exclusive interactions. The distributions are normalized to unity.
}
\label{SM-hadr-phi}
\end{figure}

\begin{figure}
\centering
\includegraphics[width=0.49\textwidth]{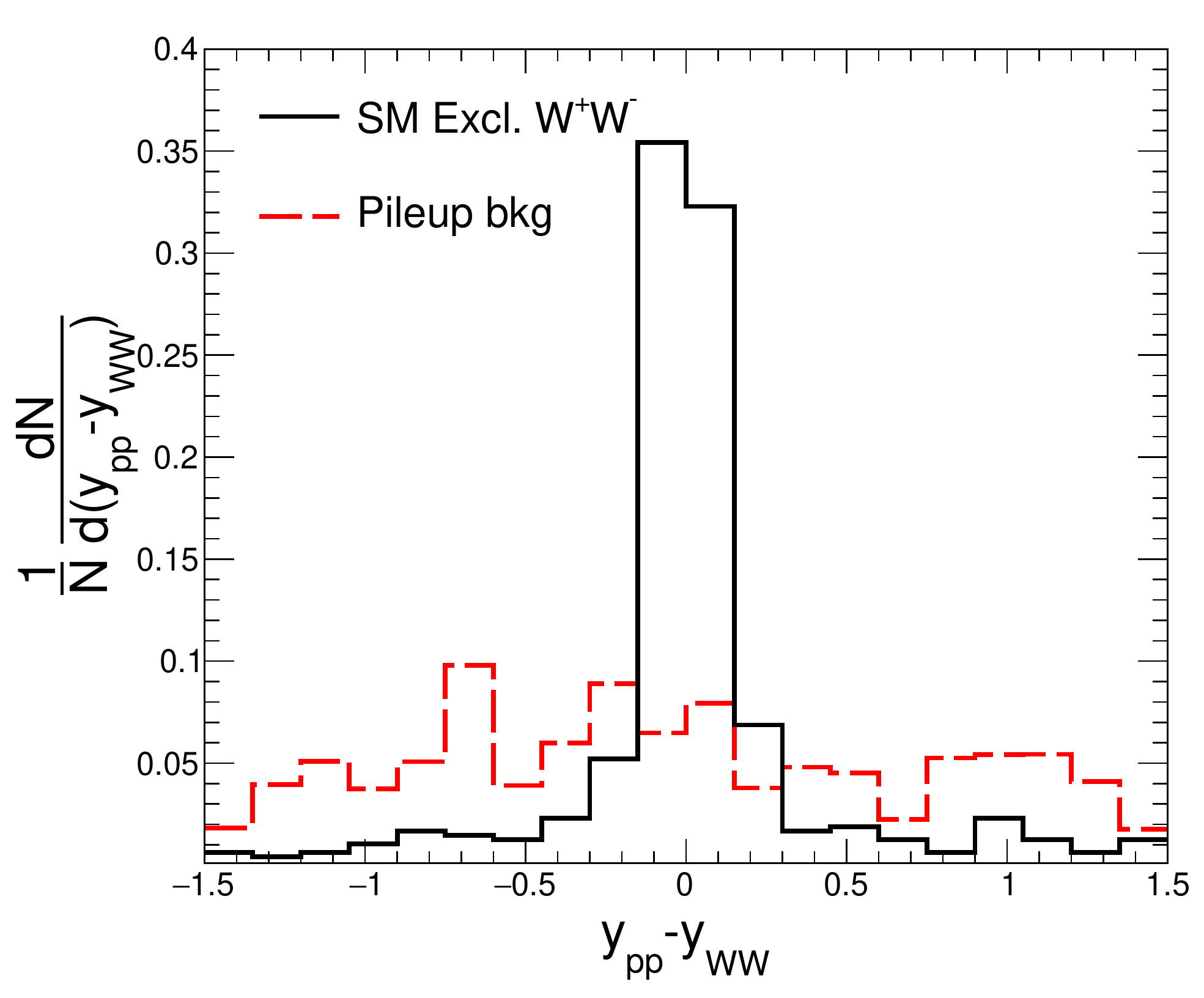}
\includegraphics[width=0.49\textwidth]{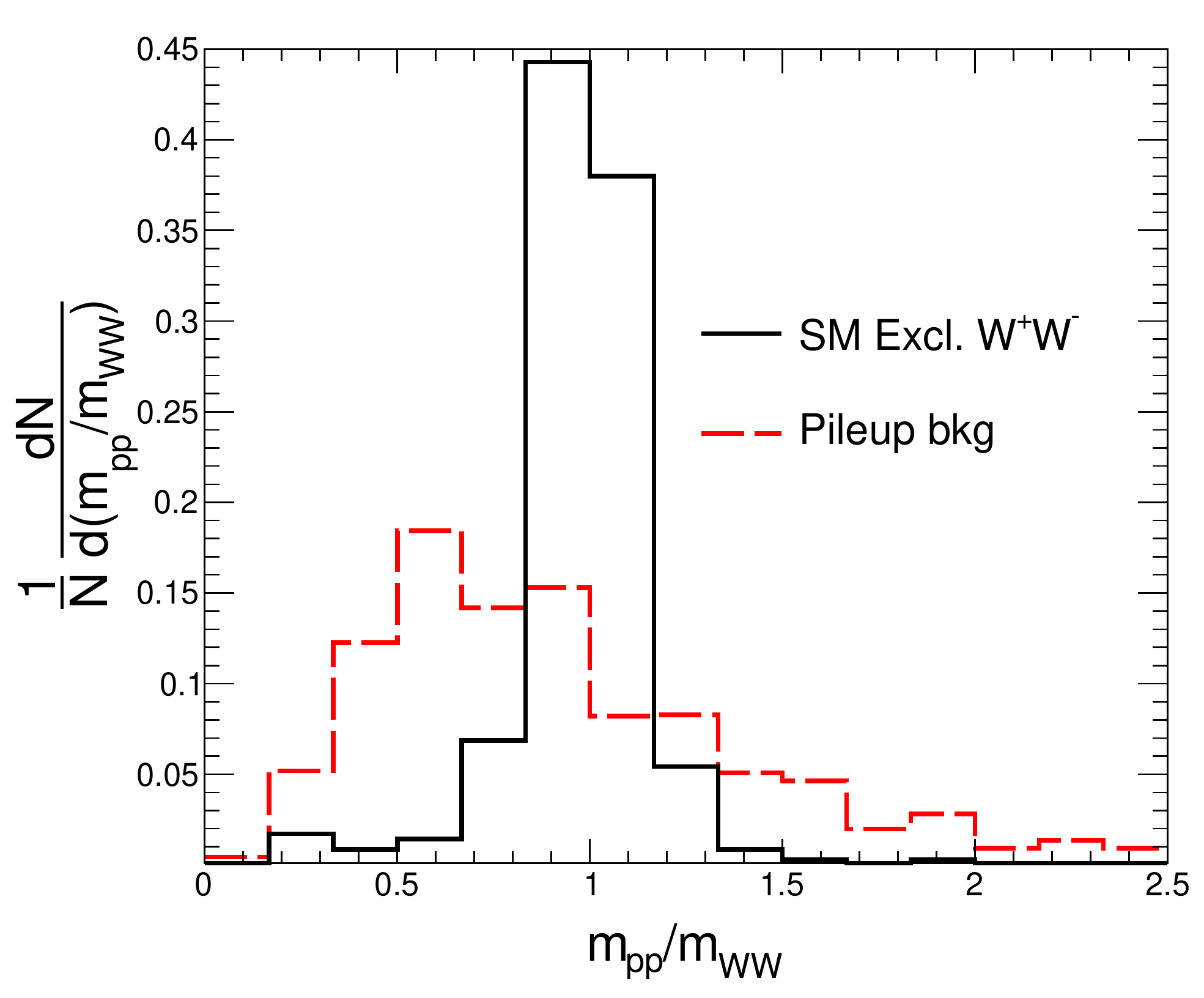}
\caption{Distributions of difference in rapidity $y_{WW} - y_{pp}$ (left) and mass ratio $m_{pp}/m_{WW}$ (right) between the $W^+W^-$ and $pp$ system for central exclusive $W^+W^-$ (black solid line) and pileup background (red dashed line) to illustrate the effectiveness of the mass and rapidity matching in order to reject pileup background. The figures are plotted after the $m_\text{j1}$, $m_\text{j2}$, $m_{WW}$, $p_T^\text{j2}/p_T^\text{j1}$ and $|\phi_\text{j1}-\phi_\text{j2}|$ selection requirements described in text. The distributions are normalized to unity.}
\label{SM-hadr-rap}
\end{figure}

\section{Study of standard model $\gamma \gamma \rightarrow W^+W^-$ events}\label{sec:analysis_strategy}

\subsection{Event selection}

Our focus in this section is to describe the event selection of high-mass $W^+W^-$ exclusive production within the SM. As mentioned in the previous section, we are mostly interested in suppressing the pileup background since this will be the leading one. It is clear that requiring two intact protons to be detected in PPS or AFP is not enough, since protons originating from pileup might overlap with a non-exclusive event. 

For pileup protons, we use a distribution $1/\xi$ (as expected for soft diffractive reactions), and consider the nominal acceptance of the proton detectors to be $0.015<\xi<0.15$. The protons originating from the additional interactions are kept if they are detected within the RP acceptance. We assume that the number of pileup interactions follow a Poisson distribution with mean 50, which is a conservative expectation for the mean number of interactions per bunch crossing at the CERN LHC during Run-3. For each pileup event, we evaluate the probability of having a proton tag for protons coming from soft diffractive events from these pileup interactions. The probability of detecting protons within the RP acceptance is calculated with Pythia8 simulated events~\cite{pythia8}, as described in Ref.~\cite{timing_detectors_saimpert}. This is the same approach adopted in Refs.~\cite{Sylvain_PRD,yyyy_Fichet,yyyZ_Baldenegro,Baldenegro_ALP}. If more than one proton is detected on each side, we select the proton pair which has the smallest $y_{pp}$ since, for most cases, the $W^+W^-$ are produced centrally in the detector at a rapidity of 0, with about 68\% of the events within   
0.8 units following the FPMC results. About 15\% of events have more than two protons in the RP acceptance from pileup interactions. The two protons are required to originate from the same vertex as the exclusive $W^+W^-$ decay products by measuring the protons time-of-flight difference with a 50 or 20 ps precision. It allows us to have a further suppression of the pileup background by up to a factor of about 40 for an average pileup of 50 interactions per bunch crossing if the timing precision is of the order of 20 ps. This selection is the same for all possible decays of the $W$ bosons.

In the next subsections, we discuss the specific event selection requirements optimized for the hadronic (two large-radius jet, zero isolated leptons), semi-leptonic (one large-radius jet, one isolated lepton) and leptonic (two isolated leptons, no jets) channels, where a lepton is considered to be isolated if the distance between it and all other particles originating from the same vertex $\Delta R_{lp}=\sqrt{(\Delta y_{lp})^2 + (\Delta \phi_{lp})^2}$ is larger than 0.4. The selection requirements used in this phenomenology study yield mutually exclusive event categories.
 
\subsection{Hadronic decays of both $W$ bosons}\label{subsec:hadronic}

We consider the hadronic decays of both $W$ bosons, leading to the production of at least two large-radius jets with $p_T^\text{jet}>100$ GeV, and no isolated leptons with $p_T^\ell> 25$ GeV. The latter condition ensures the exclusivity of the $W^+W^-$ signal event category.

Not all of the $W$ boson jets result in single large-radius jets. Most of the time, these are partially merged, so they cannot all be tagged in a single large-radius jet. In order to account for these partially merged jet topologies, we combine nearby large-radius jets with $p_T>25$ GeV if they are separated by $\Delta R_{jj} = \sqrt{\Delta\phi_{jj}^2+\Delta y_{jj}^2} < 2$ with respect to the hardest $p_T^\text{jet}$ candidates, and calculate the vector sum of the four-momenta of these two nearby jets. If they yield a better invariant mass on the reconstructed $W$ boson jet candidate, the event is retained, and the jets are merged. If the softer large-radius jets do not yield a better $W$ boson invariant mass (i.e., yields a $m_j$ closer to $m_W = 80.379$ GeV), or if they are farther than $\Delta R_{jj} > 2$, the event is rejected. In this paper, we refer to this procedure as ``jet merging''. For SM $\gamma\gamma\rightarrow W^+W^-$, the latter constitutes about 45\% of events in the fully-hadronic final state (it occurs about 25\% of the semi-leptonic cases). Events with exactly two large-radius jets or two partially-merged jets are kept after this requirement. This set of requirements is referred to as ``preselection'' in the rest of the paper.

Each of the reconstructed large-radius jets must have an invariant mass of $70 < m_\text{j1}<90$ GeV and $60 <m_\text{j2}<85$ GeV for the leading and second leading jet in mass, respectively (see Fig.~\ref{SM-hadr-mass}). 
In order to favor exclusive $W^+W^-$ events and suppress pileup background, we require the two large-radius jets to be back-to-back in the transverse plane since the $W$ bosons recoil against each other since there is no additional radiation in the process. In practice, we consider only large-radius jets strictly back-to-back $|\Delta \phi - \pi|<0.01$ rad. As shown in Fig.~\ref{SM-hadr-phi}, this selection requirement largely suppresses the pileup background, while the central exclusive production is retained. This is a particularly important selection requirement, as it exploits the $2\rightarrow 2$ topology of the signal process. Event candidates are required to have $m_{WW} > 500$ GeV. This is to further suppress non-diffractive backgrounds, which dominate at lower invariant masses $m_{WW}$. Since exclusive $W$ boson are balanced in the transverse plane, we further require them to satisfy $p_\text{T}^\text{j2}/p_\text{T}^\text{j1}>0.90$. Similar requirements were found to optimize the signal-to-background ratio in exclusive $\gamma\gamma$ production and exclusive $Z\gamma$ production phenomenology studies~\cite{yyyy_Fichet,yyyZ_Baldenegro}.

In order to suppress the remaining pileup background, we exploit the kinematic correlation between the intact protons and the reconstructed $W^+W^-$ system for exclusive $W^+W^-$ production. The mass and rapidity computed using the $W^+W^-$ bosons are, within uncertainties, similar to the ones computed using  $\xi_1$ and $\xi_2$ of the protons, The distributions of $y_{pp}-y_{WW}$ and $m_{pp}/m_{WW}$ are shown in Figs.~\ref{SM-hadr-rap}. In Figs.~\ref{SM-hadr-rap}, signal events are observed to peak at $y_{pp}-y_{WW}=0$ and $m_{pp}/m_{WW}=1$ due to the aforementioned kinematic correlation between protons and the $W^+W^-$ system. The distributions of pileup background events in these variables are respectively found to be rather flat for $y_{pp}-y_{WW}$ or decreasing for $m_{pp}/m_{WW}$ in the hadronic channel, since the protons are uncorrelated to the central $W^+W^-$ boson pair. We select events that satisfy $|y_{pp}-y_{WW}|<0.2$ and $0.7<m_{pp}/m_{WW}<1.6$ simultaneously. This selection requirement strongly suppresses the remaining pileup backgrounds, which in turn allows us to extract the exclusive $W^+W^-$ contributions. A summary of the number of events assuming a luminosity of 300 fb$^{-1}$ at $\sqrt{s} = 14$ TeV is given in Table~\ref{SM-hadr}. We assume, in addition, a precision of the fast timing detectors of either 50 or 20 ps in order to measure the proton time-of-flight. We predict a $W^+W^-$ exclusive signal of 92 events for a background of about 1640 events for 300 fb$^{-1}$. The QCD jet component dominates by far the pileup background in the fully-hadronic channel.

For this phenomenology study, we have used the invariant mass of the jets to separate quark and gluon jets from quantum chromodynamics interactions from the $W$ boson jets created in central exclusive production, as this is one of the most robust jet substructure variables~\cite{jetSubstructureReview}. The bottom two rows in Table~\ref{SM-hadr} represent an estimation of possible further suppression of the remaining QCD jet background in association with pileup protons based on the use of $N$-subjettiness ratio $\tau_2/\tau_1$ cuts, which quantifies the likelihood that the large-radius jet has a two-prong substructure, as first suggested in Ref.~\cite{Thaler:2010tr}, further confirmed by the CMS and ATLAS experiments~\cite{WbosonBoostedCMS,WbosonBoostedATLAS,WbosonBoostedATLAS2}. Based on these performance reports, it was shown that the $N$-subjettiness ratio cut $\tau_2/\tau_1 < 0.5$ can improve the QCD jet background rejection by an additional factor of $\approx 6$ in addition to the jet mass selection requirement. The jet mass requirement used in our analysis, together with a $\tau_2/\tau_1$ requirement, yield a combined inefficiency of about 20\% for $W$ boson jets decay relative to the yields extracted with the jet mass cut alone, while the background rejection factor from the jet mass requirement together with the ratio $\tau_2 /\tau_1$ cut increases by a factor of about 6 when tagging both $W$ boson jets, relative to the jet mass cut alone~\cite{WbosonBoostedCMS}. This estimation assumes that the jet mass requirement and the $\tau_2/\tau_1$ cut is mostly independent from the jet kinematics, which seems to be true for most of the $p_T$ values of interest as reported by the CMS in Ref~\cite{WbosonBoostedCMS}. Thus, the 1600 events at 20 ps time-of-flight precision could be further reduced to about 270 with the additional use of $N$-subjettiness ratio, with a decrease in the signal efficiency of about 80\% relative to the jet mass requirement alone used in our study. A combination of very precise time-of-flight measurement, together with a judicious choice of jet substructure variables, could help cope with the large QCD jets background, leading to a signal-to-background ratio of about 25\%.

Recently, the ATLAS Collaboration presented a new set of tools based on the so-called Unified Flow Object (UFO) algorithm~\cite{largeRjetATLAS}, which showed a significant improvement on the background rejection rate of QCD jets while retaining a good signal efficiency rate for retaining $W$ boson jets. Thus, the hadronic channel can become quite competitive in the study of high mass central exclusive $W^+W^-$ production, if it is supplemented by these jet substructure techniques developed by the ATLAS and CMS Collaborations. This calls for a judicious selection of jet substructure variables or use of advanced techniques by the experimental collaborations. For a thorough review on the performance of these methods, which are outside of the scope of this paper, we refer to Ref.~\cite{jetSubstructureReview}.

\begin{table}

\renewcommand{\arraystretch}{1.2}
  \makebox[\textwidth]{
\resizebox{1.3\textwidth}{!}{%
\centering
\begin{tabular}{c||c||c|c|c|c|c|c|c|c||c}
\hline
Selection                           & Excl. & $W^+W^-$ & $WZ$ & $ZZ$ & $Wj$ & $Zj$ & $t\bar{t}$ & single-top & QCD jets & Total \\

requirements                           & $W^+W^-$ & +pileup & pileup & +pileup & +pileup & +pileup & +pileup & +pileup & +pileup &  bkg. \\ \hline\hline
Preselection                                     & 990            & 1.4$\times 10^{5}$    & 7.8$\times 10^{4}$    & 2.7$\times 10^{4}$    & 1.9$\times 10^{7}$    & 7.7$\times 10^{6}$    & 1.1$\times 10^{6}$          & 5.2$\times 10^{5}$    & 1.4$\times 10^{10}$          & 1.4$\times 10^{10}$  \\ \hline
$70 < m_\text{j1} < 90$ GeV                      & 591            & 4.1$\times 10^{4}$    & 1.4$\times 10^{4}$    & 6$\times 10^{3}$    & 3.8$\times 10^{6}$    & 1.1$\times 10^{6}$    & 2.04$\times 10^{5}$          & 9.30$\times 10^{4}$    & 8.3$\times 10^{8}$          & 8.34$\times 10^{8}$  \\ \hline
$65 < m_\text{j2} < 85$ GeV                      & 274            & 1.5$\times 10^{4}$    & 4$\times 10^{3}$    & 2.1$\times 10^{3}$    & 2.5$\times 10^{5}$    & 6.8$\times 10^{4}$    & 6.35 $\times 10^{4}$          & 4.72$\times 10^{3}$    & 4.62$\times 10^{7}$          & 4.7$\times 10^{7}$  \\ \hline
|$\Delta\phi_{j1j2}-\pi$|\textless{}0.01         & 203            & 2.2$\times 10^{3}$    & 521         & 265         & 2.4$\times 10^{4}$    & 5.1$\times 10^{3}$    & 2.4$\times 10^{3}$          & 237         & 3.62$\times 10^{6}$          & 3.65$\times 10^{6}$  \\ \hline
$m_{WW} > 500$ GeV                               & 143            & 1.09$\times 10^{3}$    & 264         & 151         & 1.9$\times 10^{4}$    & 4.2$\times 10^{3}$    & 1.2$\times 10^{3}$          & 139         & 2.43$\times 10^{6}$          & 2.45$\times 10^{6}$  \\ \hline
$p_\text{T}^\text{j2}/p_\text{T}^\text{j1}>0.90$ & 142            & 1.04$\times 10^{3}$    & 221         & 135         & 1.0$\times 10^{4}$    & 2.48$\times 10^{3}$    & 708               & 65          & 1.43$\times 10^{6}$          & 1.45$\times 10^{6}$  \\ \hline
$|y_{pp}-y_{WW}| < 0.2$                          & 100            & 182         & 31          & 27          & 1112        & 201         & 28                & 11          & 1.51$\times 10^{5}$          & 1.52$\times 10^{5}$  \\ \hline
$0.7 < m_{pp}/m_{WW} < 1.6$                      & 95             & 92          & 15          & 13          & 589         & 87          & 0                 & 7           & 6.45$\times 10^{4}$          & 6.53$\times 10^{4}$  \\ \hline
$\delta t  = 50$ ps & 92 & 12 & 0 & 1 & 34  & 0 & 0 & 0 & 3.8$\times$10$^3$ & 3.9$\times$10$^3$ \\
$\delta t = 20$ ps                               & 92             & 6          & 0           & 0           & 15          & 2           & 0                 & 0           & 1.62$\times 10^{3}$          & 1.64$\times 10^{3}$  \\
\hline\hline

\hline\hline
$\delta t  = 50$ ps + & 69 & 9 & 0 & 0 & 29  & 0 & 0 & 0 & 633 & 670 \\
$\tau_2/\tau_1$ estimation & &  &  &  &   &  &  &  &  &  \\
$\delta t = 20$ ps +                         & 69             & 0          & 0           & 0           & 3          & 0           & 0                 & 0           &  270            & 273  \\
$\tau_2/\tau_1$ estimation & &  &  &  &   &  &  &  &  &  \\
\end{tabular}
}
}
\caption{Number of events for 300 fb$^{-1}$ at 14 TeV after each selection criterion for $W^+W^-$ exclusive signal (when both $W$s decays hadronically into large-radius jets), and non-diffractive $W^+W^-+$ pileup, $W^\pm Z+$ pileup, $ZZ+$ pileup, $W+$jet backgrounds. The preselection requires the presence of two large-radius jets of $p_T > 100$ GeV each, no isolated lepton and at least one proton on each side with $0.015 < \xi < 0.15$. The four first lines describe the selection on the $W$ bosons sides using the central CMS or ATLAS detector, the two next lines the exclusivity requirements using the proton detectors. Yields estimated for time-of-flight difference with precision values of 50 ps or 20 ps are shown as well. The bottom rows correspond to an estimation based on the $N$-subjettiness ratio $\tau_2/\tau_1$ cut based on ATLAS and CMS performance results~\cite{WbosonBoostedCMS,WbosonBoostedATLAS,WbosonBoostedATLAS2}.
}
\label{SM-hadr}
\end{table}

\begin{figure}
\centering

\includegraphics[width=0.49\textwidth]{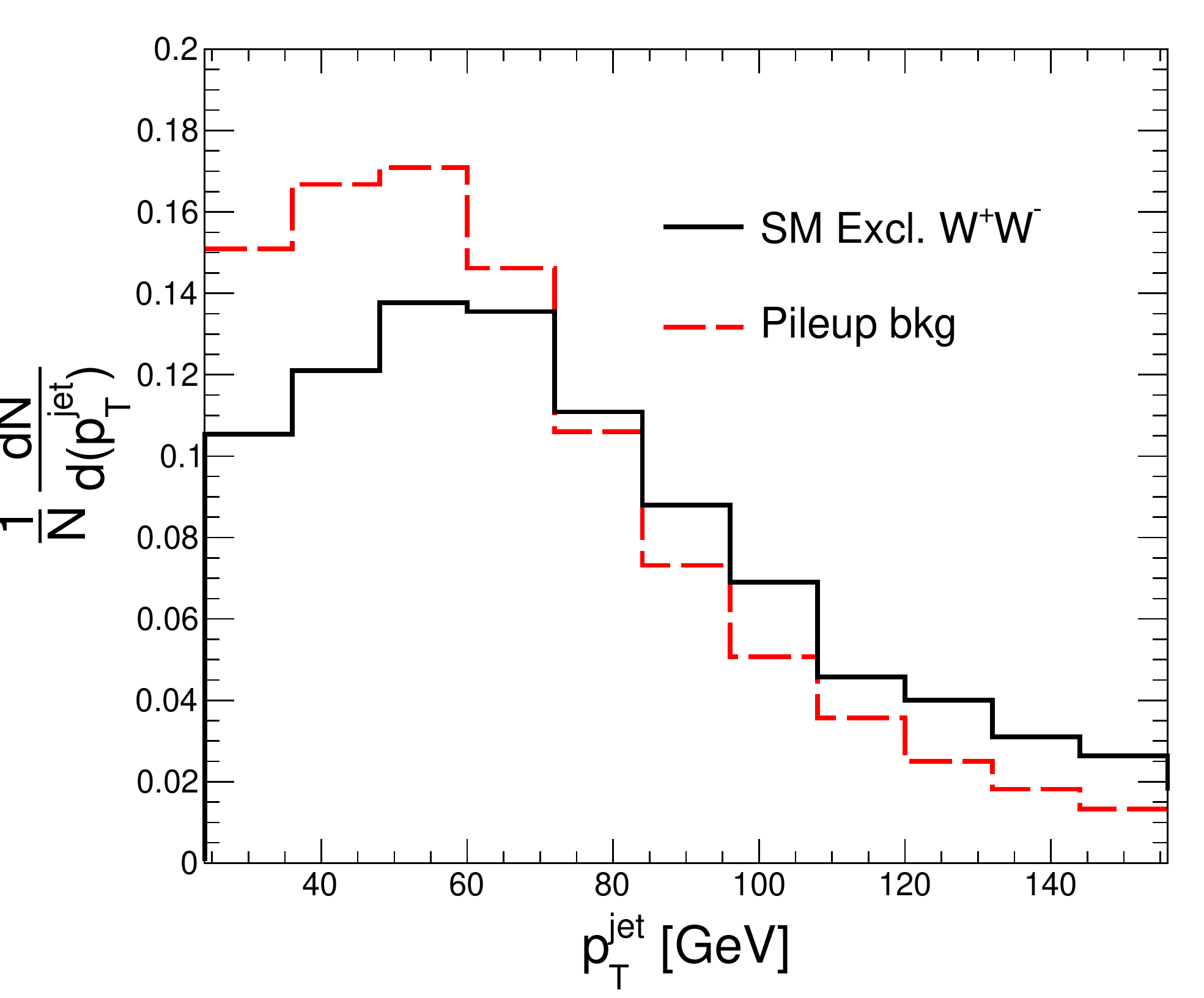}
\includegraphics[width=0.49\textwidth]{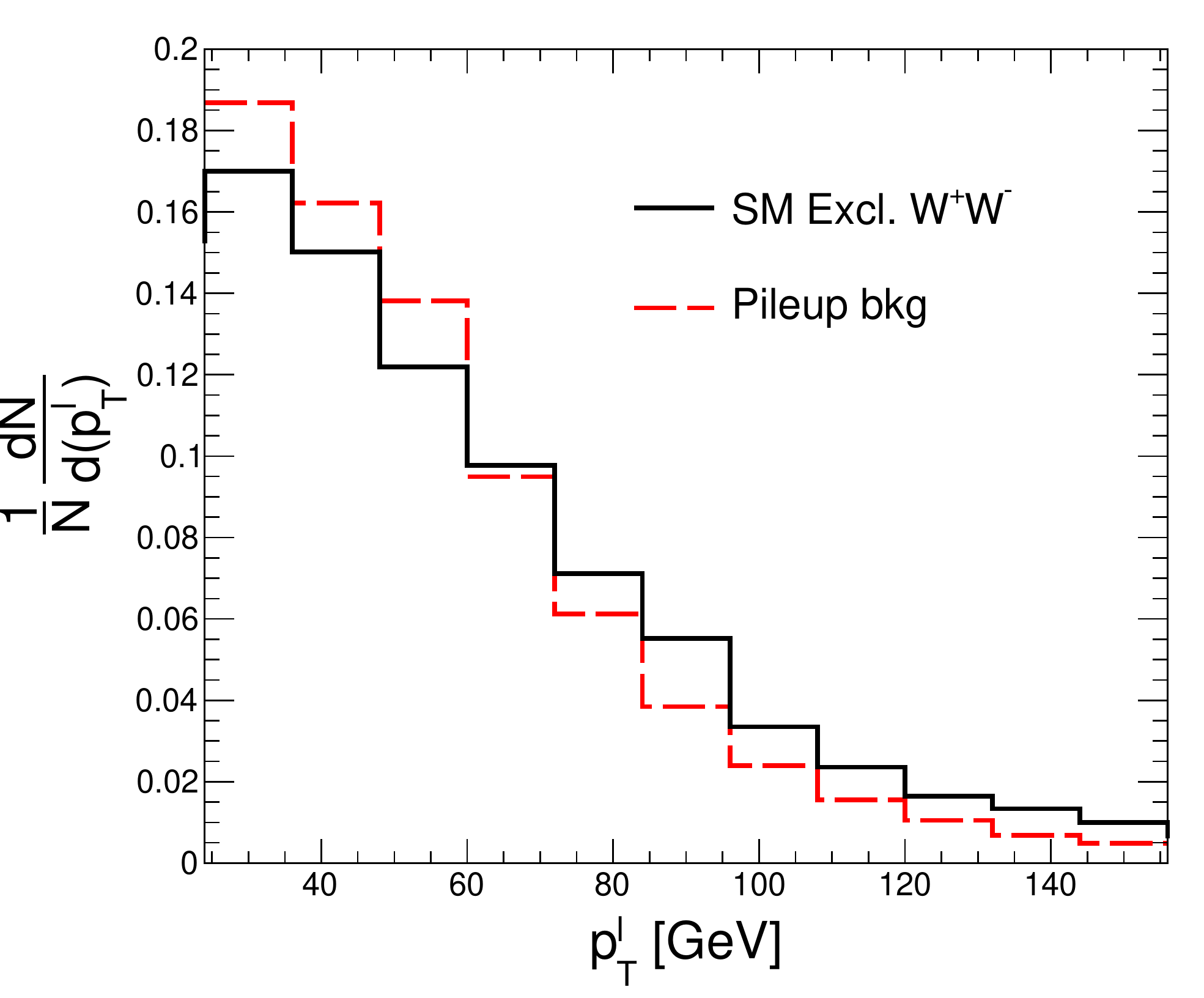}
\caption{Jet $p_T$ (left) and charged lepton $p_T$ (right) distributions normalized to unity for semi-leptonic SM exclusive $W^+W^-$ production (black full line) and pileup background (red dashed line) after preselection requirements. }
\label{SM-sl-pt}
\end{figure}

\begin{figure}
\centering
\includegraphics[width=0.49\textwidth]{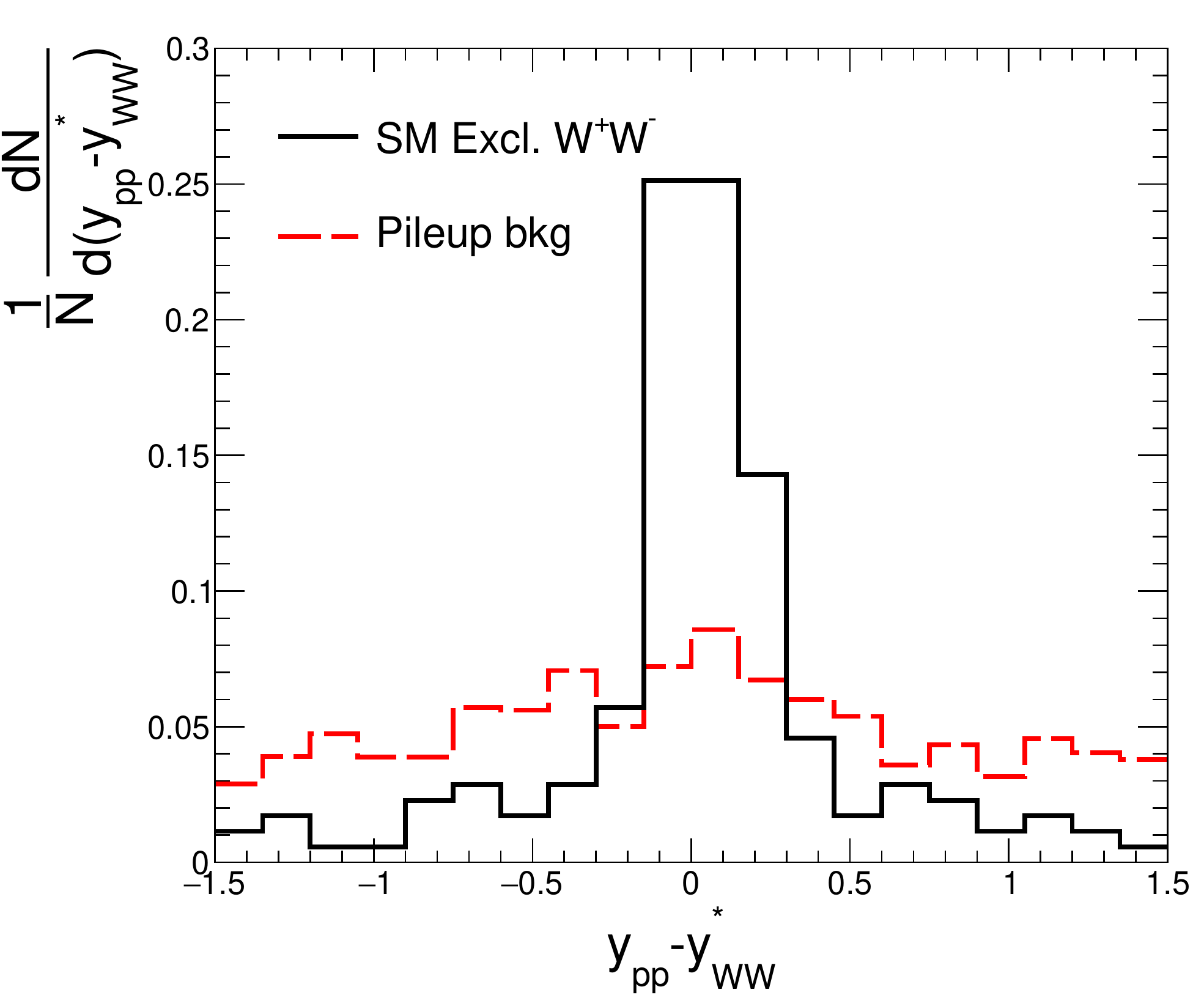}
\includegraphics[width=0.49\textwidth]{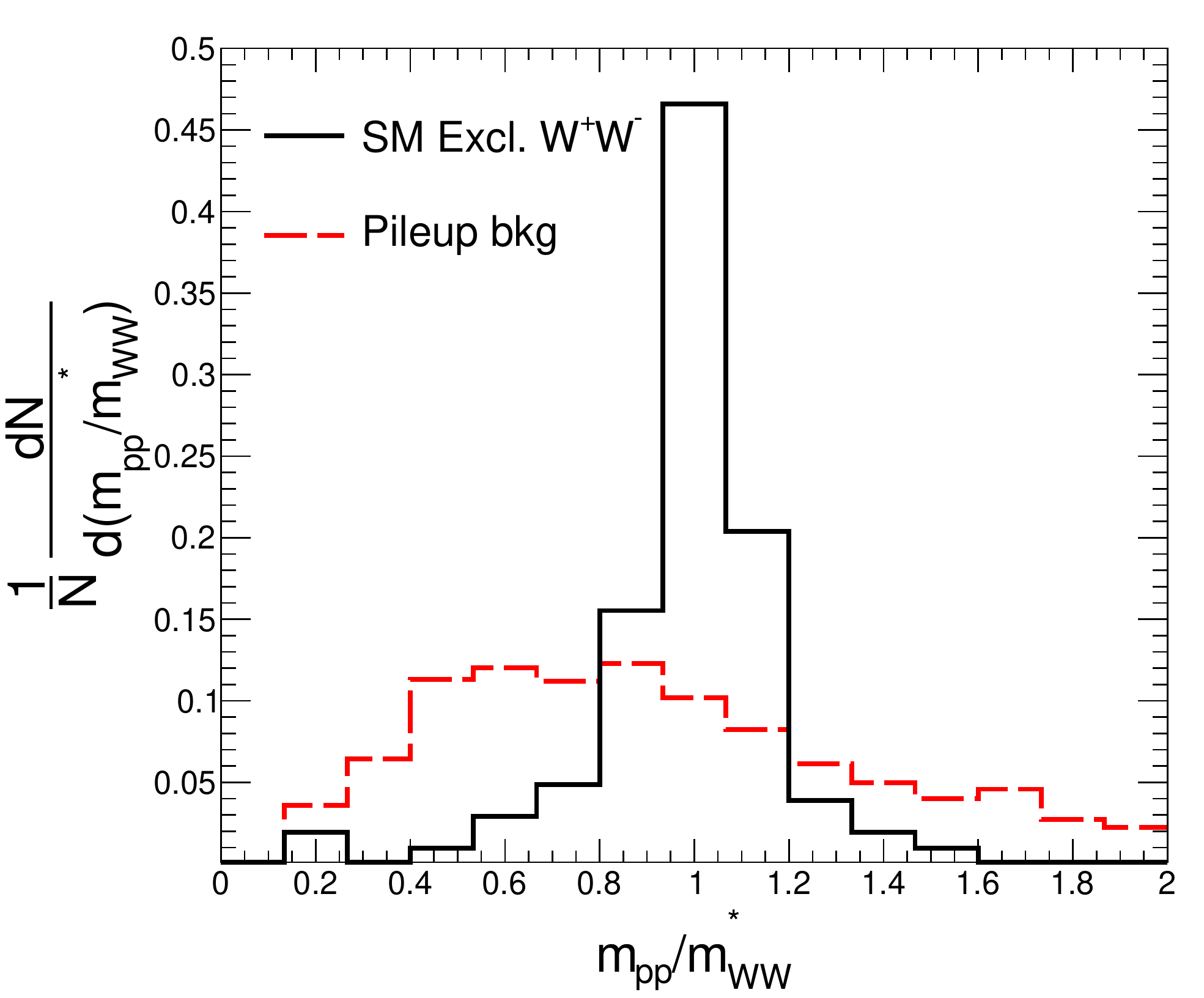}
\caption{Rapidity difference $y_{pp}-y^*_{WW}$ (left) and mass ratio $m_{pp}/m^*_{WW}$ (right) distributions for semi-leptonic decays of SM exclusive $W^+W^-$ events (black full line) and pileup background (red dashed line) normalized to unity, after preselection requirements, and lepton jet kinematic requirements summarized in Table~\ref{SM-sl}.
}
\label{SM-sl-rap}
\end{figure}

\subsection{Semi-leptonic decays of the $W$ bosons}

\begin{center}
\begin{table}[]
\makebox[\textwidth]{
\renewcommand{\arraystretch}{1.2}
\resizebox{1.2\textwidth}{!}{%

\begin{tabular}{c||c||c|c|c|c|c|c|c||c}
\hline
Selection            & Excl.  & $W^+W^-$ & $W^\pm Z$ & $ZZ$ & $Wj$ & $Z j$ & $t\bar{t}$    & single-top       & Total \\

requirements            &  $W^+W^-$ & +pileup & +pileup & +pileup & +pileup & +pileup & +pileup   & +pileup       &  bkg. \\ \hline\hline
Preselection                      & 3.9$\times 10^{3}$ & 6$\times 10^{5}$  & 7.3$\times 10^{4}$  & 5.9$\times 10^{4}$  & 6.8$\times 10^{6}$    & 5.3$\times 10^{5}$    & 1.3$\times 10^{5}$ & 3.2$\times 10^{4}$ & 8.3$\times 10^{6}$  \\ 
$p_\text{T}^\text{j1} > 120$ GeV  & 7.6$\times 10^{2}$ & 5.8$\times 10^{4}$  & 8.4$\times 10^{3}$  & 7.1$\times 10^{3}$  & 3.6$\times 10^{6}$    & 2.6$\times 10^{5}$    & 4.4$\times 10^{4}$ & 1.9$\times 10^{4}$ & 4$\times 10^{6}$  \\ 
$p_\text{T}^\ell > 100$ GeV       & 334      & 2.3$\times 10^{4}$  & 3014      & 2.52$\times 10^{3}$  & 1.56$\times 10^{6}$    & 1.92$\times 10^{5}$    & 1.5$\times 10^{4}$ & 5979     & 1.8$\times 10^{6}$  \\ 
$m^{*}_{WW} > 500$ GeV            & 144      & 7698      & 1040      & 1077      & 70595       & 7970        & 699      & 987      & 9$\times 10^{4}$  \\ \hline
$70 < m_\text{j1} < 90$ GeV       & 74       & 2035      & 171       & 216       & 4211        & 459         & 224      & 14       & 7331      \\ 
$p_\text{T}^\text{miss} > 30$ GeV & 52       & 1628      & 141       & 172       & 3540        & 242         & 196      & 0        & 5919      \\ \hline
$|y_{pp}-y^*_{WW}| < 0.2$           & 29       & 265       & 31        & 31        & 587         & 28          & 84       & 0        & 1026      \\
$0.7 < m_{pp}/m^{*}_{WW} < 1.3$   & 27       & 101       & 13        & 16        & 168         & 16          & 28       & 0        & 342       \\ \hline
$\delta t = 50$ ps & 22 & 6 & 2 & 0.5 & 10 & 1 & 2 & 0 & 21.5 \\
$\delta t = 20$ ps                & 22       & 5         & 1         & 0         & 4           & 0           & 1        & 0        & 11        \\ \hline
\end{tabular}}}
\caption{Number of events for 300 fb$^{-1}$ after each selection criterion for $W^+W^-$ exclusive signal (when one $W$ boson decays leptonically and the other one into hadrons), non-diffractive $W^+W^-+$ pileup, $W^\pm Z+$ pileup, $ZZ+$ pileup, $W^\pm+$jet +pileup, $t\bar{t}$+pileup, single-top+pileup backgrounds. The bottom two rows correspond to the options of time-of-flight difference measurement of the scattered protons with precision of 50 ps or 20 ps.}
\label{SM-sl}
\end{table}
\end{center}

Turning to the semi-leptonic decay scenario, we start by requiring the presence of at least one large-radius jet with $p_T^\text{jet}> 25$ GeV and exactly one charged isolated lepton of $p_T^\ell>25$ GeV. As in the previous section, if there is more than one large-radius jet, we merge jets within $\Delta R_{jj} < 2$ with the criteria described in Sec.~\ref{subsec:hadronic}, then exactly one large-radius jet or merged dijet with $p_T^\text{j} > 120$ GeV. The slightly larger $p_T^\text{jet}$ is optimized to yield a better signal-to-background ratio. This is the respective ``preselection'' step for this channel.

Unlike the hadronic decay channel, a complete reconstruction of the final state is not possible in the semi-leptonic decay channel because of the undetected neutrino. The energy and momentum of the neutrino along the beam direction, $E^\nu$ and $p_z^\nu$, are unknown. In order to estimate the invariant mass and rapidity of the $W^+W^-$ system (which is later used to calculate the correlations between the central system and the forward protons), we consider $E^\nu = p_T^\text{miss}$ and $p_z^\nu=0$ strictly for the $m_{WW}$ and $y_{WW}$ evaluation, and assign the missing transverse momentum of the neutrino with $p_T^\text{miss}$, as is usually done. The choice of fixing $p_z^\nu = 0$ is based on the $p_z^\nu$ distribution observed in simulated events, which peaks around $0$ with a standard deviation of about $100$ GeV for central exclusive $W^+W^-$ events. This spread of $100$ GeV results in an additional smearing on the estimated invariant mass and rapidity of the reconstructed $W^+W^-$ kinematics. Since we are not reconstructing all the decay products of the $W^+W^-$ system due to the undetected neutrino, we refer to the mass and rapidity of the diboson event candidate as $m^*_{WW}$ and $y^{*}_{WW}$, respectively, to emphasize that this is an estimation of the actual variables $m_{WW}$ and $y_{WW}$ that we aim to ideally reconstruct. The $m^{*}_{WW}$ and $y^{*}_{WW}$ variables were comparable with the $y_{WW}$ and $m_{WW}$ variables at generator-level. We found that, on average, $y^{*}_{WW}$ is closer to $0$ than the true $y_{WW}$, while $m^{*}_{WW}$ is lower than $m_{WW}$ on average, due to the missing $E^\nu$ and $p_Z^\nu$ information. By comparing this estimation with truth-level information available from the MC generated event (i.e., assuming we could indeed reconstruct the neutrino $E^\nu$ and $p_z^\nu$), we found that the exclusive $W^+W^-$ signal yield would improve by a factor of about $1.6$ relative to that obtained with our conservative approach. The $E^\nu$ and $p_{z}^\nu$ of the neutrino could be estimated, up to a two-fold ambiguity, by assuming that the undetected neutrino, together with the detected charged lepton, has to yield the rest mass of the $W$ boson. Such a method was employed by the CDF Collaboration in Ref.~\cite{CDF_neutrino}. Such an approach could be considered for the experimental analysis for Run-3 at the CERN LHC.

In order to suppress the non-diffractive backgrounds, and to favor boosted topologies for large-radius jet reconstruction, we further require that $p_{T}^\text{jet} > 120$ GeV and $p_T^\ell > 100$ GeV. These requirements are determined based on the distributions observed in the pileup background and SM exclusive $W^+W^-$ signal, as shown in Fig. \ref{SM-sl-pt}. Furthermore, the invariant mass of the reconstructed central system is required to be $m^{*}_{WW}>$600 GeV. The aforementioned requirement further suppresses the non-diffractive background, and favors the reconstruction of boosted topologies of the $W$ boson decay products in large-radius jets. We use the reconstructed $p_T^\text{miss} > 30$ GeV to estimate the transverse momentum carried away by the undetected neutrino.

The rapidity difference between the central system and the forward two protons is taken as $|y^*_{WW}-y_{pp}|<0.2$ and the mass ratio $0.7< m_{pp}/m^*_{WW}<1.3$ for the semi-leptonic channel, as shown in Fig.~\ref{SM-sl-rap}. The latter set of criteria are optimized in order to reject as much pileup background as possible, while retaining most of the SM exclusive $W^+W^-$ events. Note that the selection window on these variables is narrower in this final-state, even if we do not have the $E^\nu$ and $p^\nu_z$ of the missing neutrino. The reason is that we do not have the same smearing effects as in the fully-hadronic final state. It is also worth to note that we can still exploit these kinematic correlations between the forward protons and central $W$ boson pair in semi-leptonic final states. The number of events for signal and background after each selection criterion is given in Table~\ref{SM-sl} assuming in addition a resolution of the fast timing detectors of either 50 or 20 ps in order to measure the proton time-of-flight. With this set of selection criterion, we expect 22 SM $\gamma\gamma\rightarrow W^+W^-$ events in the semi-leptonic channel, versus a expected background of $11$ events, which is largely dominated by $W^\pm$+jets with pileup protons. We found that the acoplanarity between the charged lepton and the large-radius jet does not provide much discrimination between pileup backgrounds and central exclusive $W^+W^-$ events. For this reason, no acoplanarity cut is applied in the semi-leptonic channel.

\subsection{Leptonic decays of $W$ bosons}\label{subsec:leptonic}

\begin{figure}
\centering
\includegraphics[width=0.49\textwidth]{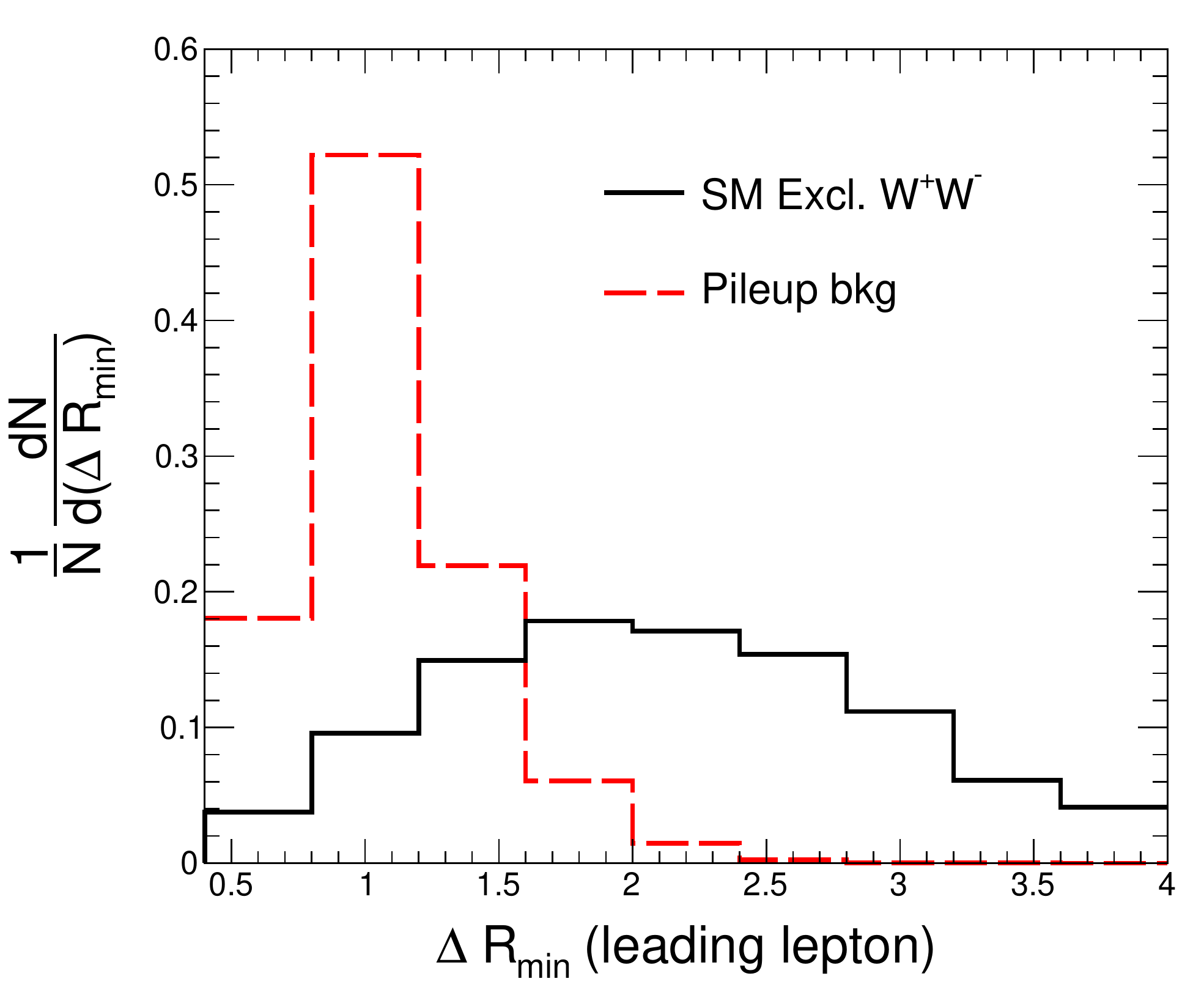}
\includegraphics[width=0.49\textwidth]{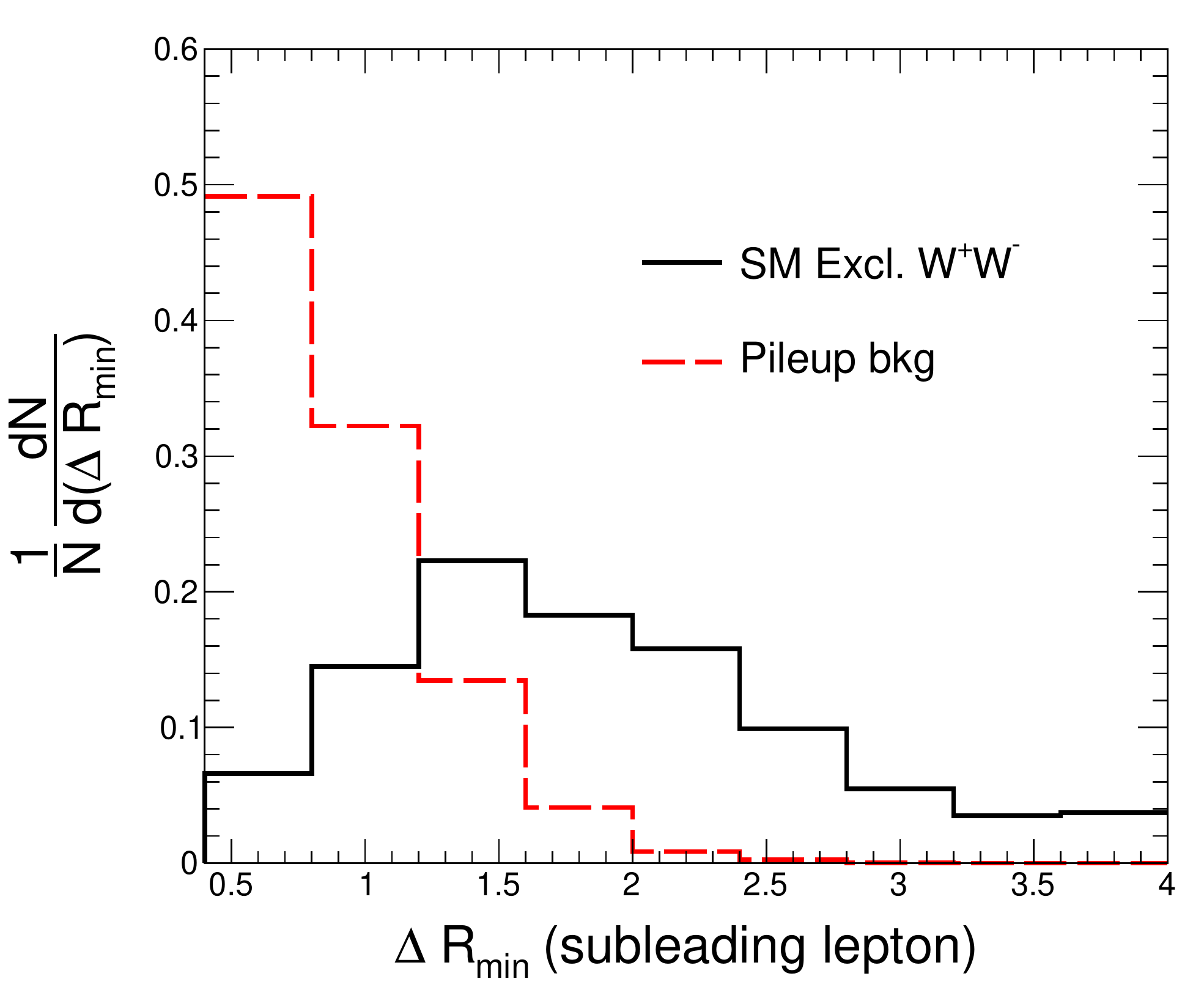}

\caption{Separation in $y$-$\phi$ space between the isolated leptons and the closest particle, $\Delta R_\text{min} = \sqrt{(\Delta y)^2+(\Delta\phi)^2}$, for exclusive $W^+W^-$ (black full line) and pileup background (red dashed line) for leading $p_T$ lepton (left), and subleading-$p_T$ when $p_T>25$ GeV, for $W^+W^-$ purely leptonic decay channel. The distributions are normalized to unity.
}
\label{SM-lept-rmin}
\end{figure}

\begin{figure}
\centering
\includegraphics[width=0.49\textwidth]{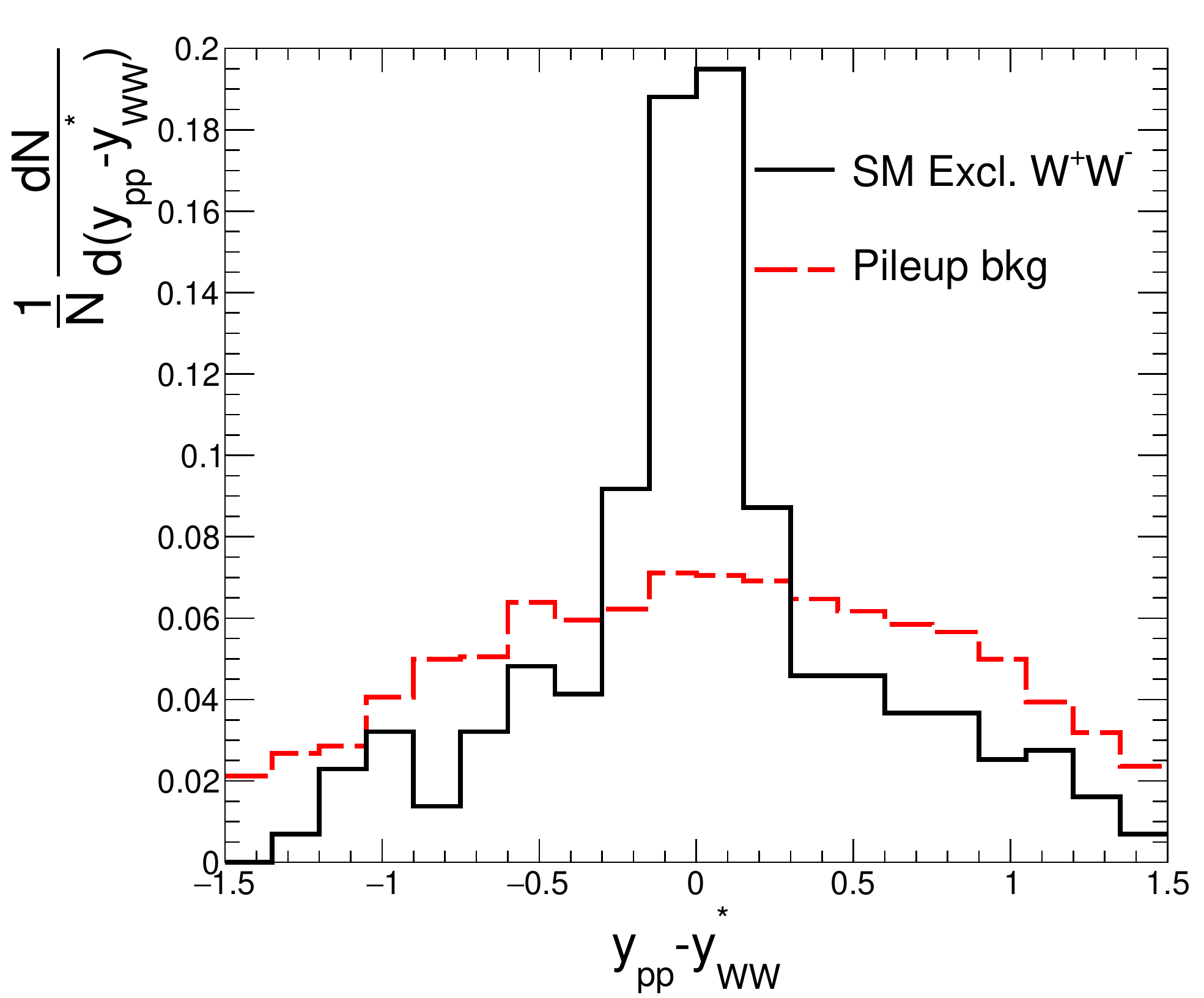}
\includegraphics[width=0.49\textwidth]{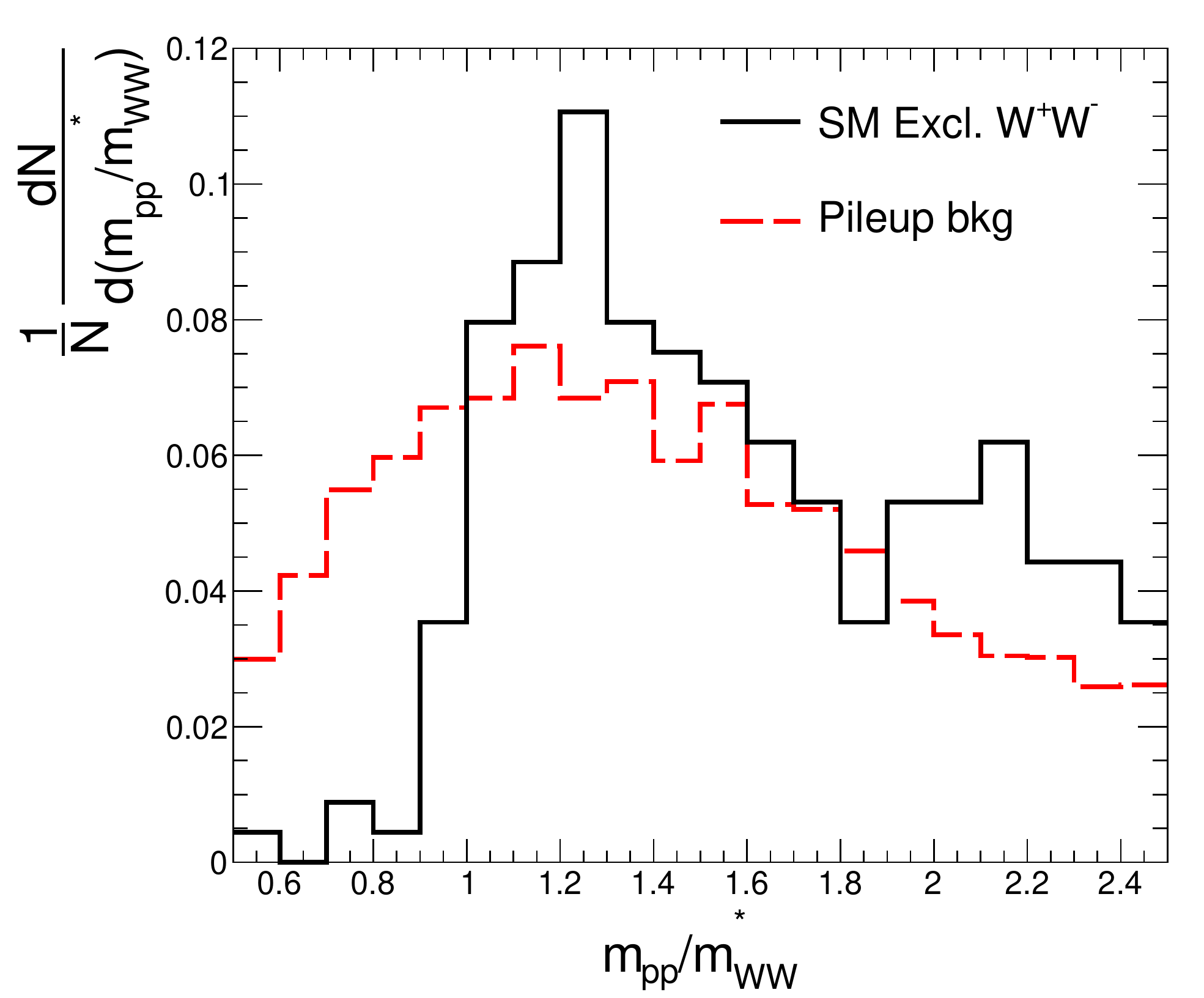}
\caption{Rapidity difference $y_{pp}-y^*_{WW}$ (left) and mass ratio $m_{pp}/m^*_{WW}$ (right) distributions for leptonic decays of SM exclusive $W^+W^-$ events (black full line) and pileup background (red dashed line) normalized to unity, after preselection requirements, and lepton kinematic requirements summarized in Table~\ref{SM-lept}.
}
\label{SM-leptonic-rap}
\end{figure}

\begin{table}
\center
\renewcommand{\arraystretch}{1.25}
\resizebox{\textwidth}{!}{%
\begin{tabular}{c|c|c|c|c|c}
\hline
Selection requirement & SM Excl. & $W^+W^-$ & $W^\pm Z$ & $ZZ$ &  Total\\
 &$W^+W^-$ & + pileup & + pileup & + pileup   & bkg.\\
\hline
Preselection & 522 & 2.29$\times$10$^4$
& 9 359 & 895 & 3.3$\times$10$^4$ \\ 
(Two leptons w/ $p_T^\ell>25$ GeV) & &  &  &  &  \\ \hline
$\Delta R_\text{min}>1.5$ & 300 & 250 & 87 & 16 & 353 \\ \hline
$p_T^{\ell 1}>50$ GeV & 188 & 91 & 57 & 9  &  126\\\hline
$p_T^\text{miss}>30$ GeV & 140 & 77 & 46 & 3 &  122 \\\hline
$\Delta \phi_{\ell\ell} <2.7$ or different flavor  & 131 & 74 & 45 & 3  &  122\\\hline
 $|y^{*}_{WW}-y_{pp}|<0.5$ & 86 & 25 & 19 & 1 &  45 \\\hline
$0.5< m_{pp}/m^{*}_{WW} < 2.5$ & 64 & 6 & 7 & 1 &  14 \\\hline
$\delta t = 50$ ps & 47 & 2 & 0 & 0 & 2\\
$\delta t = 20$ ps & 45 & 2 & 0 & 0 & 2 \\
\hline
\end{tabular}}
\caption{Number of events for 300 fb$^{-1}$ after each selection criterion for $W^+W^-$ exclusive signal (where both $W$ bosons decay leptonically), non-diffractive $W^+ W^-+$ pileup, $W^\pm Z+$ pileup, $ZZ+$ pileup in leptonic final states. The bottom two rows correspond to the options of time-of-flight difference measurement of the scattered protons with precision of 50 ps or 20 ps.}
\label{SM-lept}
\end{table}

For the leptonic case, we require at least two isolated leptons of opposite electric charge with transverse momentum $p_T^\ell > 25$ GeV each, and a veto on large-radius jets with $p_T^\text{j}>100$ GeV to suppress non-diffractive contributions, and to ensure exclusivity of event category. The jet veto requirement suppresses the contribution of non-diffractive $t\bar{t}$, single-top, $W^\pm$+jet, and $Z$+jet in association with pileup protons. The two leptons can be of different or same flavors. This is dubbed ``preselection'' for the leptonic channel.

As in the semi-leptonic channel, in order to estimate the invariant mass of the exclusive $W^+W^-$ candidate, we fix $E^\nu =  p_T^\text{miss}$ and $p_Z^\nu = 0$ for each of the undetected neutrinos in the calculation of $m_{WW}^*$ and $y_{WW}^*$, similar as in the semi-leptonic scenario. We use the reconstructed $p_T^\text{miss} > 30$ GeV to estimate the transverse momentum carried away by the undetected neutrinos. For same-flavor leptons, an additional cut of $\Delta\phi_{\ell\ell} < 2.7$ is applied to suppress contributions from $\gamma\gamma\rightarrow \ell^+\ell^-$ and Drell-Yan + pileup contributions, which are negligible after these event selection requirements. The reason is that same-flavor lepton pairs from these processes are produced back-to-back in the transverse plane, whereas if they originate from the decay of two $W$ bosons they are expected to be less correlated in the transverse plane.

For the leptonic channel, due to the absence of underlying event activity in exclusive $W^+W^-$, the lepton isolation requirement is extended even further. Here, we consider $\Delta R_{min} \equiv \sqrt{(\Delta y)^2+(\Delta \phi)^2}$ is the distance of the charged lepton to the closest particle in rapidity and azimuth fit to the same interaction vertex. For each prompt lepton, we require $\Delta R_{min}>1.5$, since this effectively reduces the contribution of pileup backgrounds while retaining most of the signal events, as shown in Fig.~\ref{SM-lept-rmin}. We further tighten the transverse momenta requirements $p_T^{\ell 1}>50$ GeV for the leading lepton, while leaving the subleading lepton at $p_T^{\ell 2}>25$ GeV. To further reduce pileup contributions, the correlation between protons and $W^+W^-$ decay daughters, namely $|y_{pp}-y^*_{WW}|<0.5$ and a loose cut of $0.5< m_{pp} /m^*_{WW}<2.5$ is used. The cut on these variables is looser, given that in this case we do not detect two neutrinos, which worsens an accurate reconstruction of $y^{*}_{WW}$ and $m^*_{WW}$, as shown in Fig.~\ref{SM-leptonic-rap}, but profitting taking advantage that the pileup background is not very large at this stage. Assuming a time-of-flight with a precision of 20 ps, we expect about 45 events from SM $\gamma\gamma\rightarrow W^+W^-$ at a background of 2 counts from non-diffractive $W^+W^-$ in association with pileup protons. The $W$ boson events in the purely leptonic channel are mostly at low $m_{WW}$ and at low $p_T^W$, distinct from the phase space covered by the hadronic and semi-leptonic channels, hence the difference in the integrated number of events.

\subsection{Discussion of standard model analysis}\label{sec:sm_results}

We note that only in the semi-leptonic and hadronic channels one can access high mass $W^+W^-$ production with large $W$ boson transverse momenta $p_T^W$ for future differential cross section studies, which are absolutely necessary to test predictions based on the SM non-abelian coupling between photons and $W$ bosons. The purely leptonic channel offers the possibility of studying the SM processes, albeit at lower masses and softer $p_T$ of the $W$ boson. For leptons with $p_T^{\ell1,\ell2} > 100$ GeV, the expected number of exclusive $W^+W^-$ events is 0. In other words, for harder $\gamma\gamma\rightarrow W^+W^-$ production can only be studied in the semi-leptonic and hadronic channels.

We emphasize that exclusive production of $W^+W^-$ boson pairs can be used as a standard candle process for a better understanding of the RP detector calibration and proton reconstruction, as well as for a better understanding of trigger efficiency studies. In fact, since every decay channel is independent from one another, they can be used separately to cross check the proton spectrometer overall calibration. This is complementary to the $pp \rightarrow p \ell^+\ell^- p$ process, where $\ell = e, \mu$, used by the CMS-TOTEM and ATLAS Collaborations to calibrate their proton spectrometers~\cite{dilepton_PPS, dilepton_AFP}. More concretely, calibration at higher $\xi$ can be achieved with the hadronic and semi-leptonic channels, while the calibration at low $\xi$ could be done with the purely leptonic channel.

\section{Anomalous $\gamma\gamma \rightarrow W^+W^-$ scattering}\label{sec:anomalous_coupling}

\subsection{Effective field theory framework}
Deviations from the SM from possible new physics contributions at energies much larger than those accessible at the LHC can be described in an effective field theory formalism. Indeed, by integrating out potential heavy degrees of freedom present at the new physics energy scale or beyond, residual interaction terms are obtained at energy scales presently probed at LHC energies~\cite{EFT1,EFT2}. This corresponds to gauge invariant non-renormalizable effective operators, which allows us to constrain several families of extensions of the SM in a single framework. In this paper, we consider dimension-six operators directly related to $\gamma\gamma W^+W^-$ quartic couplings. By imposing U(1)$_\text{em}$ and global custodial SU(2)$_\text{C}$ symmetries, two such operators are allowed with their respective coupling strength parameters denoted by $a_0^W$ and $a_C^W$. The effective field theory framework used in the present paper follows of Ref.~\cite{ggWW_ChaponKepkaRoyon,PhysRevD.78.073005}. More concretely, the following effective interaction Lagrangian is used,

\begin{equation}
\mathcal{L}_6^\text{eff} = -\frac{e^2}{8} a_0^W F_{\mu\nu}F^{\mu\nu}W^{+\alpha}W^{-}_{\alpha}-\frac{e^2}{16} a_C^W F_{\mu\alpha}F^{\mu\beta} \Big( W^{+\alpha}W^{-}_{\beta}+W^{-\alpha}W^{+}_{\beta} \Big)
\end{equation}

Where $F$ and $W$ represent the field strength tensors of electromagnetic and weak interactions after electroweak symmetry breaking, respectively. The interaction $\gamma\gamma \rightarrow W^+W^-$ induced by this effective operator may violate unitarity at high energies. In order to suppress this unphysical feature of the anomalous $\gamma\gamma\rightarrow W^+W^-$ process, we consider a dipole form factor with a cutoff scale $\Lambda_\text{cutoff}$ which modifies the coupling as:

\begin{equation}
    a_{0,C}^W (W_{\gamma\gamma}^2) \rightarrow \frac{a_{0,C}^W}{\Big(1+W_{\gamma\gamma}^2/\Lambda_\text{cutoff}^2\Big)^2}
\end{equation}

Here, $W_{\gamma\gamma}$ is the center of mass energy of the initial-state diphoton system and $\Lambda_\text{cutoff}$ represents the energy scale where new physics may manifest. For our projections, we consider a scenario where $\Lambda_\text{cutoff} = 2 $ TeV, and the case where $\Lambda_\text{cutoff} \rightarrow \infty$, which corresponds to the case where unitarity might be violated, i.e., no form factor. In our case, most of the events produced at high $W^+W^-$ invariant mass are rejected in our analysis due to the RP acceptance cut, $\xi < 0.15$. Because of this, it has been found in previous studies with intact protons that the difference in the results with and without form factors is not very large (see for example Refs.~\cite{yyyZ_Baldenegro,yyyy_Fichet,ggWW_ChaponKepkaRoyon}). A direct relation between dimension-eight and dimension-six couplings $\gamma\gamma W^+W^-$ operator is found when assuming that an anomalous $WWZ\gamma$ vertex vanishes, as discussed in Refs.~\cite{dimension8-1,dimension8-2}. Therefore, our projections can be mapped to the dimension-eight couplings under this assumption.

\subsection{Event selection}

\begin{figure}[th!]
\centering
\includegraphics[width=0.45\linewidth]{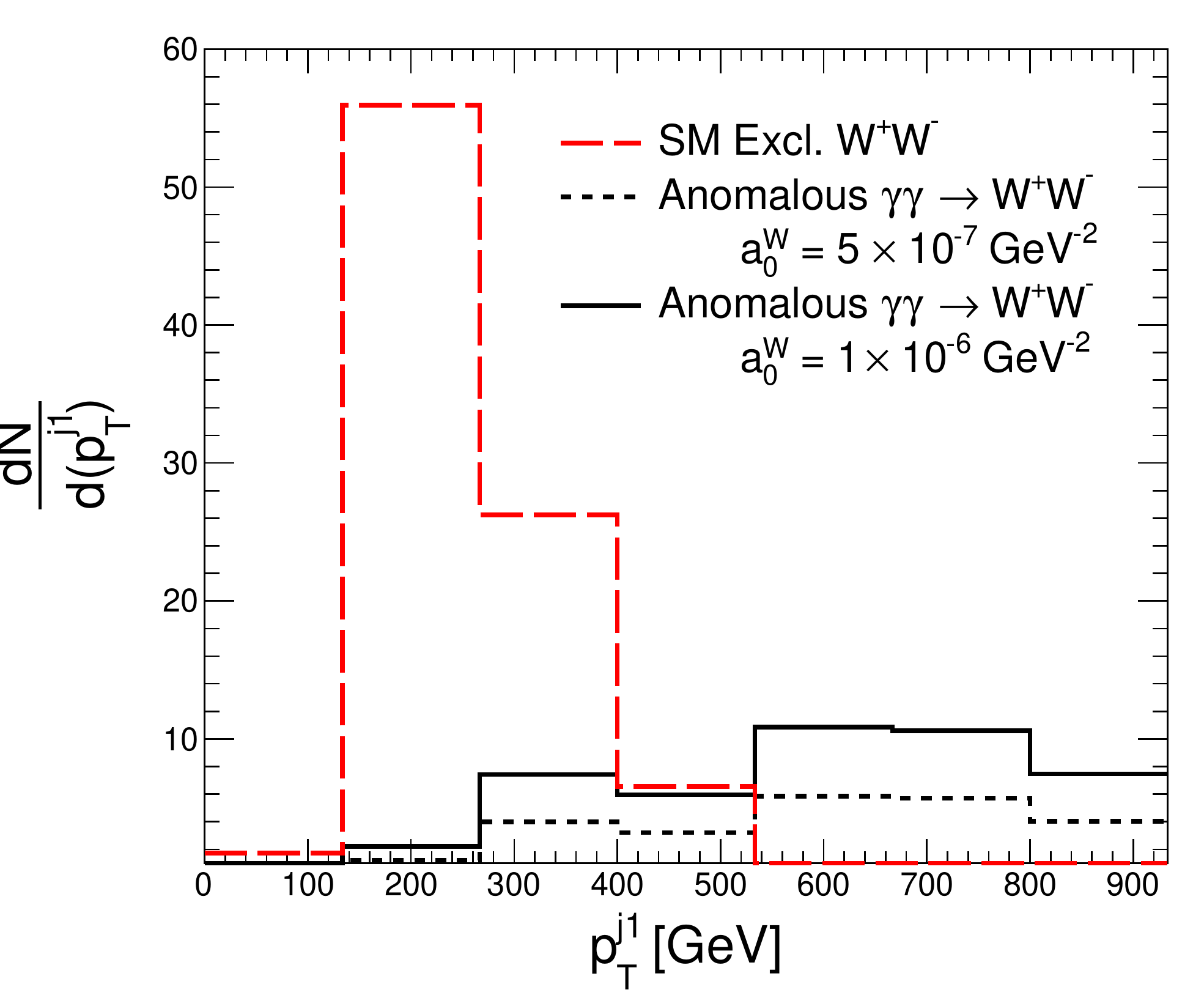}
\includegraphics[width=0.45\linewidth]{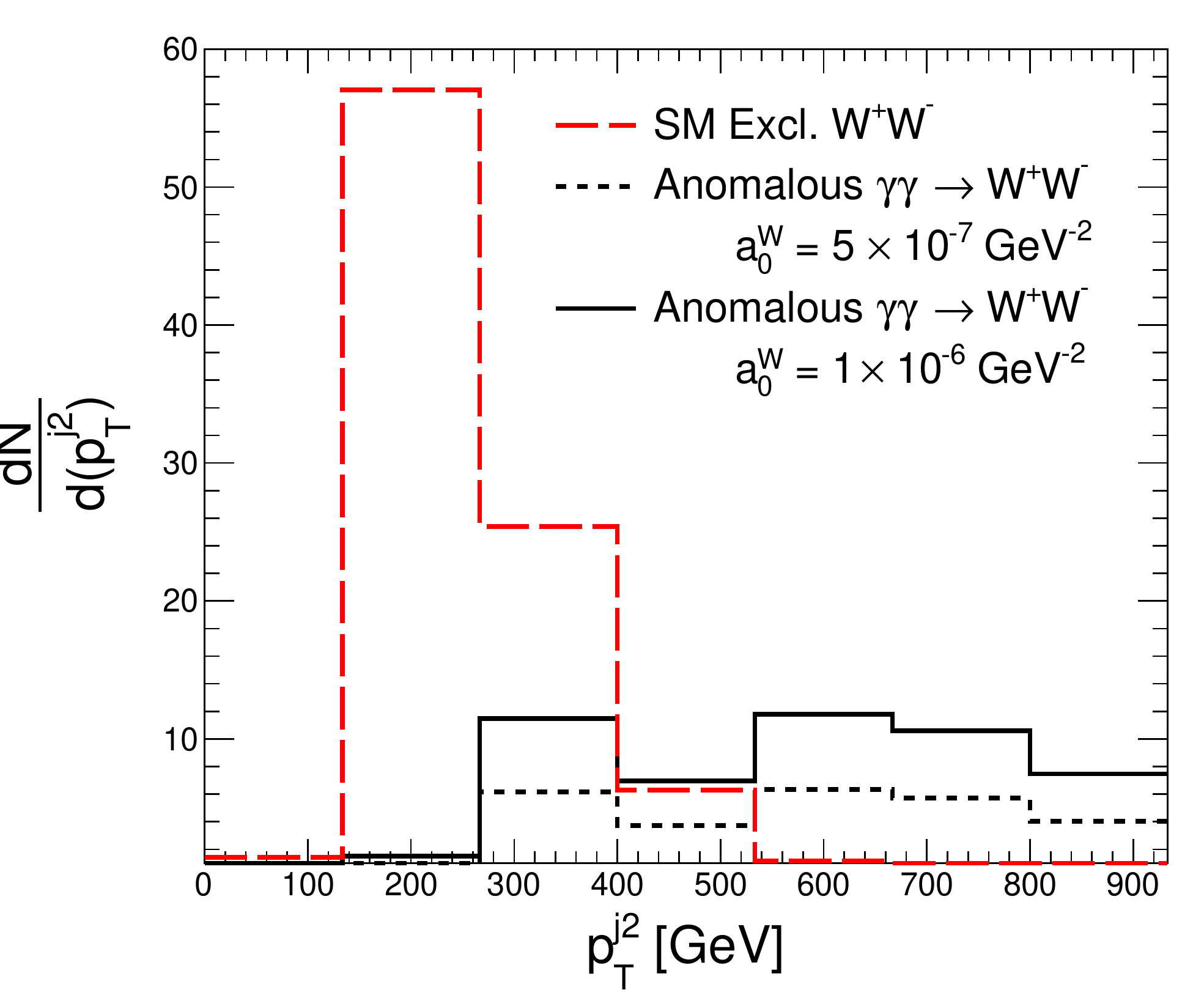}
\caption{Leading (left) and second leading (right) large-radius jet $p_T$ for the $\gamma\gamma\rightarrow W^+W^-$ process, where each $W$ boson decays hadronically. The anomalous contribution is shown in black solid and dashed lines, which represent two different values of anomalous coupling $a_0^W = 5\times 10^{-7}$ and $a_0^W = 10^{-6}$ GeV$^{-2}$ with fixed $a_C^W = 0$. The SM $\gamma\gamma\rightarrow W^+W^-$ contribution is shown in the red dashed line  (after time-of-flight requirement, normalized to 300 fb$^{-1}$ of luminosity). No matching cut in rapidity or mass is applied for these plots.}
\label{AC-hadr-pt1}
\end{figure}

\begin{figure}[tb!]
\centering
\includegraphics[width=0.7\linewidth]{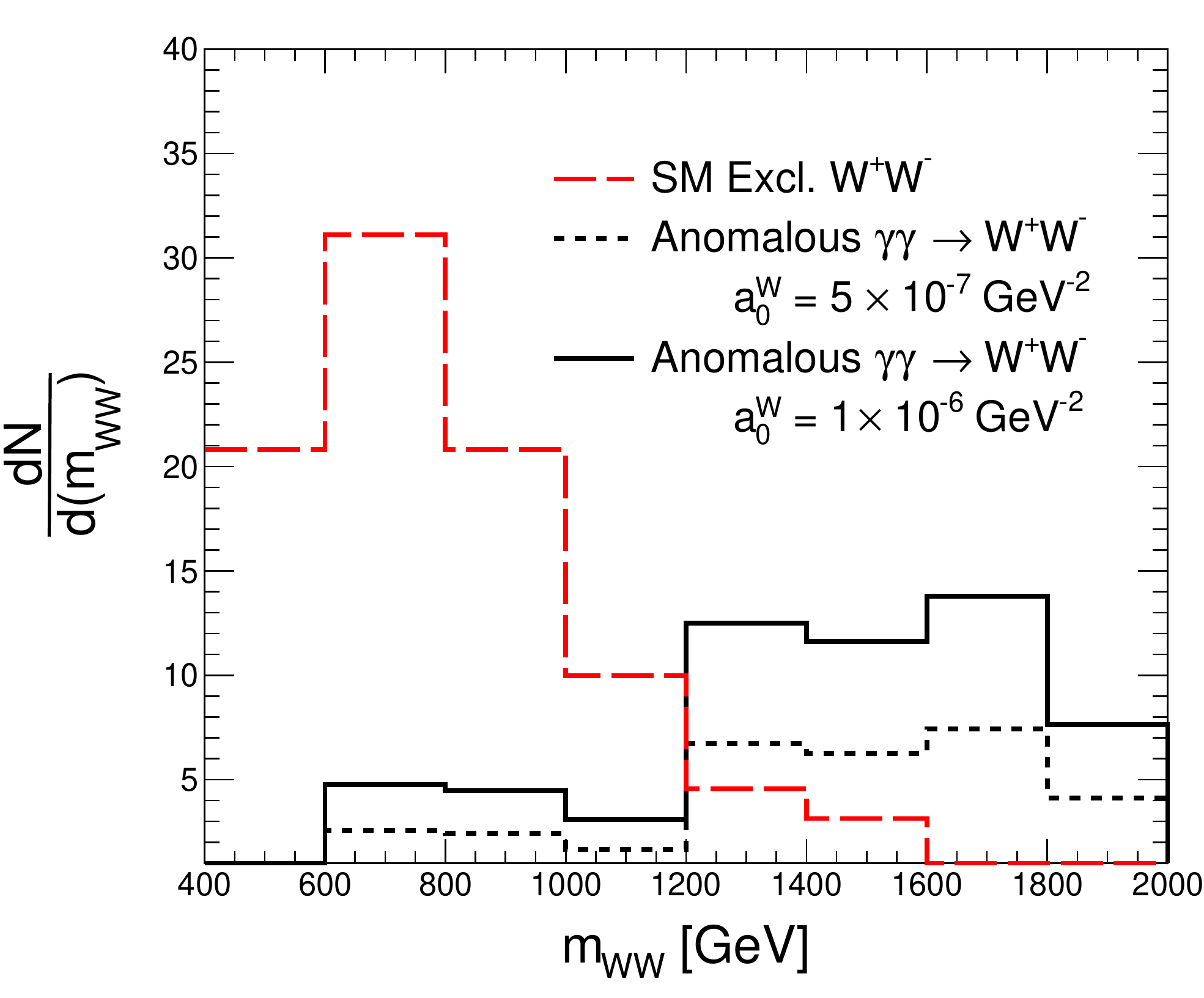}
\caption{$W^+W^-$ invariant mass reconstructed using the fully-hadronic final state. The SM $\gamma\gamma\rightarrow W^+W^-$ contribution is in red dashed line, whereas the anomalous contribution is in black solid and dashed lines, corresponding to different coupling values. The distributions are normalized to an integrated luminosity of 300 fb$^{-1}$. Contributions induced by anomalous couplings are present mostly at higher mass (no form factor applied).}
\label{AC-hadr-WWmass}
\end{figure}

In contrast to the SM $\gamma\gamma \rightarrow W^+W^-$ process, the production rate of central exclusive $W^+W^-$ boson pairs induced by anomalous couplings increases with the invariant mass of the diboson system and with the transverse momentum of each $W$ boson, $p_{T}^W$. This means that boosted topologies will be largely favored in anomalous interactions, distinct from SM expectations. Since potential events induced by anomalous coupling interactions are expected to appear mostly at large $m_{WW}$ and $p_{T}^W$, the search strategy originally tailored to isolate SM central exclusive $W^+W^-$ production needs to be modified accordingly. For illustration, we show the 
leading and second leading large-radius jet $p_T$ and the reconstructed $W^+W^-$ mass for SM and anomalous $W^+W^-$ production in the hadronic channel in Figs.~\ref{AC-hadr-pt1} and \ref{AC-hadr-WWmass}, respectively. The anomalous contribution is shown in black for different values of anomalous couplings, whereas the SM $\gamma\gamma\rightarrow W^+W^-$ contribution is shown in red dashed line. We note that events induced by anomalous interactions contribution mostly at larger jet $p_T$ and mass. 

In the hadronic case, we require the transverse momentum of both large-radius jets to be $p_T^\text{jet}>400$ GeV (as seen in Fig.~\ref{AC-hadr-pt1}). Similarly, we require that the invariant mass of the reconstructed $W^+W^-$ candidate to be $m_{WW} > 1$ TeV (see Fig.~\ref{AC-hadr-WWmass}). The latter cut reduces most of the residual SM exclusive $W^+W^-$ contributions, while retaining a large fraction of anomalous $W^+W^-$ event candidates. The results are given in Table~\ref{AC-hadr} for hadronic decays.

In a similar way, the higher $p_T$ and mass requirements eliminate most of the SM $\gamma\gamma \rightarrow W^+W^-$ in case of the semi-leptonic decays. In this case, we tighten the jet and lepton kinematics as $p_T^\ell > 250$ GeV and $p_T^\text{jet} > 350$ GeV, with $m_{WW}^* > 1200$ GeV.

For the leptonic final state, we use the same strategies as in Sec.~\ref{subsec:leptonic}, similar to those in Refs.~\cite{PPS,AFP,ggWW_ChaponKepkaRoyon}. The one-dimensional sensitivity on anomalous couplings is of the order of $4.3\times 10^{-6}$ GeV$^{-2}$ and $1\times 10^{-5}$ GeV$^{-2}$ for $a_0^W$ and $a_C^W$ at 95\% CL for the leptonic final state alone, respectively. We note here that, in contrast to the findings of the SM analysis, the boosted large-radius jet topologies are significantly favored for events with a contribution from anomalous coupling vertices. The leptonic channel is not as sensitive to deviations from the SM as the hadronic or semi-leptonic channel. This is mostly because of the much smaller branching fraction which does not allow to populate the phase space at large $m_{WW}$ and $p_T^W$, where anomalous coupling contributions are most significant.

\begin{table}
\center
\renewcommand{\arraystretch}{1.2}
\resizebox{\textwidth}{!}{%
\begin{tabular}{c|c|c|c|c|c|c}
\hline
\multirow{2}{*}{Selection requirement} & \multicolumn{2}{c|}{$a_0^W$=$5.10^{-7}$ GeV$^{-2}$} & \multicolumn{2}{c|}{$a_0^W$=$10^{-6}$ GeV$^{-2}$} & SM Excl. & Pileup \\
\cline{2-5}
 & no f.f. & f.f. & no f.f. & f.f. & $W^+W^-$ & Background \\
 \hline
Preselection & 1 892 & 61 & 7 724 & 204 & 990 & 1.4$\times 10^{10}$  \\
\hline
 $70<m_\text{j1}< 90$ GeV & \multirow{2}{*}{1 020} & \multirow{2}{*}{26} & \multirow{2}{*}{4 320} & \multirow{2}{*}{107} & \multirow{2}{*}{274} & \multirow{2}{*}{4.7$\times 10^7$} \\
$65 <m_\text{j2}<85$ GeV & & & & & & \\ 
\hline
$|\Delta \phi_\text{j1j2} - \pi|<0.01$ & 1 010 & 25 & 4 269 & 103 & 203 & 3.7$\times 10^6$  \\
$m_{WW}>500$ GeV & 1008 & 25 & 4157 & 93 & 143 & 2.5$\times 10^6$ \\
$p_T^{j2}/p_T^{j1}>0.9$ & 978 & 24 & 4150 & 91 & 142 & 1.4$\times 10^6 $\\
\hline
 $|y_{WW}-y_{pp}|<0.2$ & 337 & 15 & 1 419 & 45 & 100 & 1.5$\times 10^5$ \\
 \hline
$0.7< m_{pp}/m_{WW}<1.6$ & 44 & 11 & 171 & 34 & 95 & 6.5$\times 10^4$ \\
\hline
$\delta t = 50$ ps & 39 & 11 & 144 & 31 & 92 & 3.9$\times 10^3$ \\
$\delta t = 20$ ps & 38 & 11 & 140 & 30 & 92 & 1.64$\times 10^3$ \\
\hline
$p_T^\text{j1,j2}>$400 GeV & \multirow{2}{*}{21.6} & \multirow{2}{*}{3.9} & \multirow{2}{*}{110.1} & \multirow{2}{*}{18.2} & \multirow{2}{*}{4.9} & \multirow{2}{*}{87} \\
$m_{WW}>$1000 GeV  &  &  &  &  &  &  \\
\hline
\end{tabular}
}
\caption{Number of events at 300 fb$^{-1}$ after each sequential selection criterion for anomalous exclusive $W^+W^-$ production with and without form factors ($a_0^W$=$5\times10^{-7}$ GeV$^{-2}$ and $a_0^W$=$10^{-6}$ GeV$^{-2}$, and when both $W$s decay hadronically), SM $W^+W^-$ exclusive production, and pileup backgrounds, i.e., the sum of non-diffractive production of $W^+W^-$, $WZ$, $ZZ$, $W^\pm+$jets, $t\bar{t}$, single-top, QCD jets with pileup protons. The bottom row corresponds to the event selection criterion optimized for the anomalous coupling analysis.}
\label{AC-hadr}
\end{table}

\begin{table}[h]
\center
\renewcommand{\arraystretch}{1.2}
\resizebox{\textwidth}{!}{%
\begin{tabular}{c||c|c|c|c|c|c}
\hline
\multirow{2}{*}{Selection requirement} & \multicolumn{2}{c|}{$a_0^W$=$5\times10^{-7}$ GeV$^{-2}$} & \multicolumn{2}{c|}{$a_0^W$=$10^{-6}$ GeV$^{-2}$} & Exclusive & Pileup \\
\cline{2-5}
 & no f.f. & f.f. & no f.f. & f.f. & $W^+W^-$ & background \\
 \hline
Preselection & 190 & 29 & 680 & 108 & 3 919 & 8.3$\times 10^6$ \\
\hline
$p_{T}^{\text{jet}}>120$ GeV & \multirow{3}{*}{90} & \multirow{3}{*}{6} & \multirow{3}{*}{346} & \multirow{3}{*}{26} & \multirow{3}{*}{144} & \multirow{3}{*}{9$\times 10^4$}\\
$p_T^\ell>100$ GeV & & & & & &\\
$m^{*}_{WW}>600$ GeV & \multirow{2}{*}{} & \multirow{2}{*}{} & \multirow{2}{*}{} & \multirow{2}{*}{} & \multirow{2}{*}{} & \multirow{2}{*}{} \\
\hline
 $70 <m_\text{j1}< 90$ GeV & 59 & 2 & 220 & 17 & 74 & 7331 \\
 \hline
 $p_T^\text{miss}>30$ GeV & 49 & 5 & 170 & 14 & 52 & 5919\\
\hline
$|y^{*}_{WW}-y_{pp}|<0.2$  & 25 & 4 & 74 & 10 & 29 & 1026 \\
 \hline
$0.7< m_{pp}/m^{*}_{WW}<1.3$ & 10 & 3 & 24 & 7 & 27 & 342 \\
\hline
$\delta t = 50$ ps & 9 & 3 & 24 & 7 & 22 & 21.5 \\
$\delta t = 20$ ps & 9 & 3 & 24 & 7 & 22 & 11 \\
\hline
$p_{T}^\text{jet}>350$ GeV & \multirow{3}{*}{1.1} & \multirow{3}{*}{0.1} & \multirow{3}{*}{8} & \multirow{3}{*}{1.1} & \multirow{3}{*}{0.9} & \multirow{3}{*}{0}\\
$p_T^\ell>250$ GeV &  &  &  &  &  & \\
$m^{*}_{WW}>1200$ GeV  &  &  &  &  &  &\\
\hline
\end{tabular}}
\caption{Number of events for an integrated luminosity of 300 fb$^{-1}$ after each selection criterion in the search of anomalous exclusive $W^+W^-$ boson pair production with and without form factors ($a_0^W$=$5\times10^{-7}$ GeV$^{-2}$ and $a_0^W$=$10^{-6}$ GeV$^{-2}$ with $a_C^W = 0$ in semi-leptonic channel. The second-to-last column corresponds to SM exclusive $W^+W^-$ boson production, and the pileup background, which includes the combination of the pileup backgrounds. The bottom row corresponds to the selection requirements optimized for the anomalous coupling studies, as described in the text.}
\label{AC-sl}
\end{table}

\subsection{Results of anomalous coupling studies}\label{sec:anomalous_results}

The expected 95\% CL limit and 5$\sigma$ sensitivity for $a_C^W$ and $a_0^W$ anomalous couplings for 300 fb$^{-1}$ of luminosity with and without form factor are shown in Figs.~\ref{fig:a0_limit_noff} and \ref{fig:a0_limit_ff}, respectively, and the one-dimensional projections on the anomalous couplings are presented in Table~\ref{final}. The statistical significance is calculated with $Z = \sqrt{2(S+B)\ln(1+S/B)-2S}$, with $S$ and $B$ representing the signal and background event counts, respectively~\cite{statistics}. By combining all the decay channels, we obtain a sensitivity of $3.7 \times 10^{-7}$ GeV$^{-2}$ for $a_0^W$ and $3\times10^{-6}$ GeV$^{-2}$ for $a_C^W$ at 95\%CL, when fixing one of the couplings to $0$ and varying the other, without the use of form factor. The hadronic channel has the best sensitivity for deviations treated in the dimension-six effective field theory formalism, as shown explicitly in the expected limit calculation in Table~\ref{AC-sl}. In order to draw a comparison with the hadronic, semi-leptonic, and leptonic final state results, we show the expected limits of the leptonic decay using similar selection criteria as in Ref.~\cite{ggWW_ChaponKepkaRoyon} in Table~\ref{final}. The sensitivity in the leptonic channel alone is similar for anomalous couplings with and without form factor, since form factor acts at larger invariant masses and $p_T$, which the leptonic channel does not populate completely. For these projections, we do not assume possible improvements with jet substructure variable cuts. We estimate that the one-dimensional bounds could be potentially be further improved down to values of $|a_0^W| < 3 \times 10^{-7}$ GeV$^{-2}$ and  $|a_C^W| < 7.4 \times 10^{-7}$ GeV$^{-2}$ at 95\% CL by applying a similar cut on the $N$-subjettiness ratio $\tau_2/\tau_1 < 0.5$, as discussed in Section~\ref{subsec:hadronic}, i.e., an improvement of factor of $\approx 1.3$ over our projections without the use of these advanced techniques. This means that, in searching for these anomalous contributions, jet substructure techniques are not as crucial as in the SM analysis.

{\begin{figure}[ht!]
\centering
\includegraphics[width=0.7\linewidth]{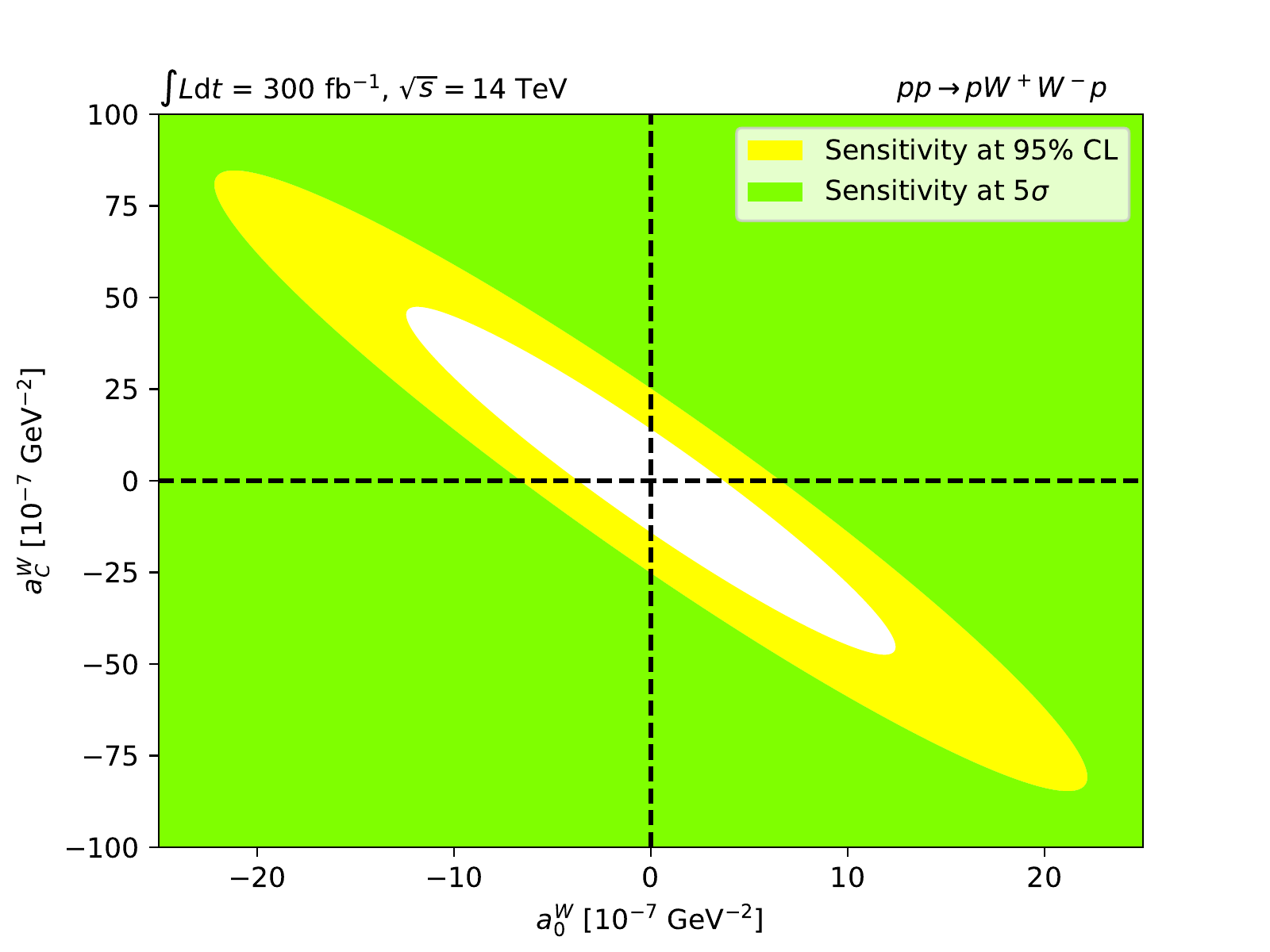}

\caption{\label{fig:a0_limit_noff} Projected sensitivities on the anomalous coupling parameters $a_0^W$ and $a_C^W$ without form factors. The projections are shown for $pp\rightarrow p W^+W^- p$ at 14 TeV assuming an integrated luminosity of 300 fb$^{-1}$. The yellow and green areas represent respectively the projected sensitivities at 95\% CL and $5\sigma$ combining the hadronic, semi-leptonic, and leptonic decay channels of the $W^+W^-$ system. The blank area in the center represents the region where we do not expect sensitivity to the anomalous coupling parameter. Time-of-flight measurements with 20 ps precision is assumed.}
\end{figure}

\begin{figure}[hb!]
\centering
\includegraphics[width=.7\linewidth]{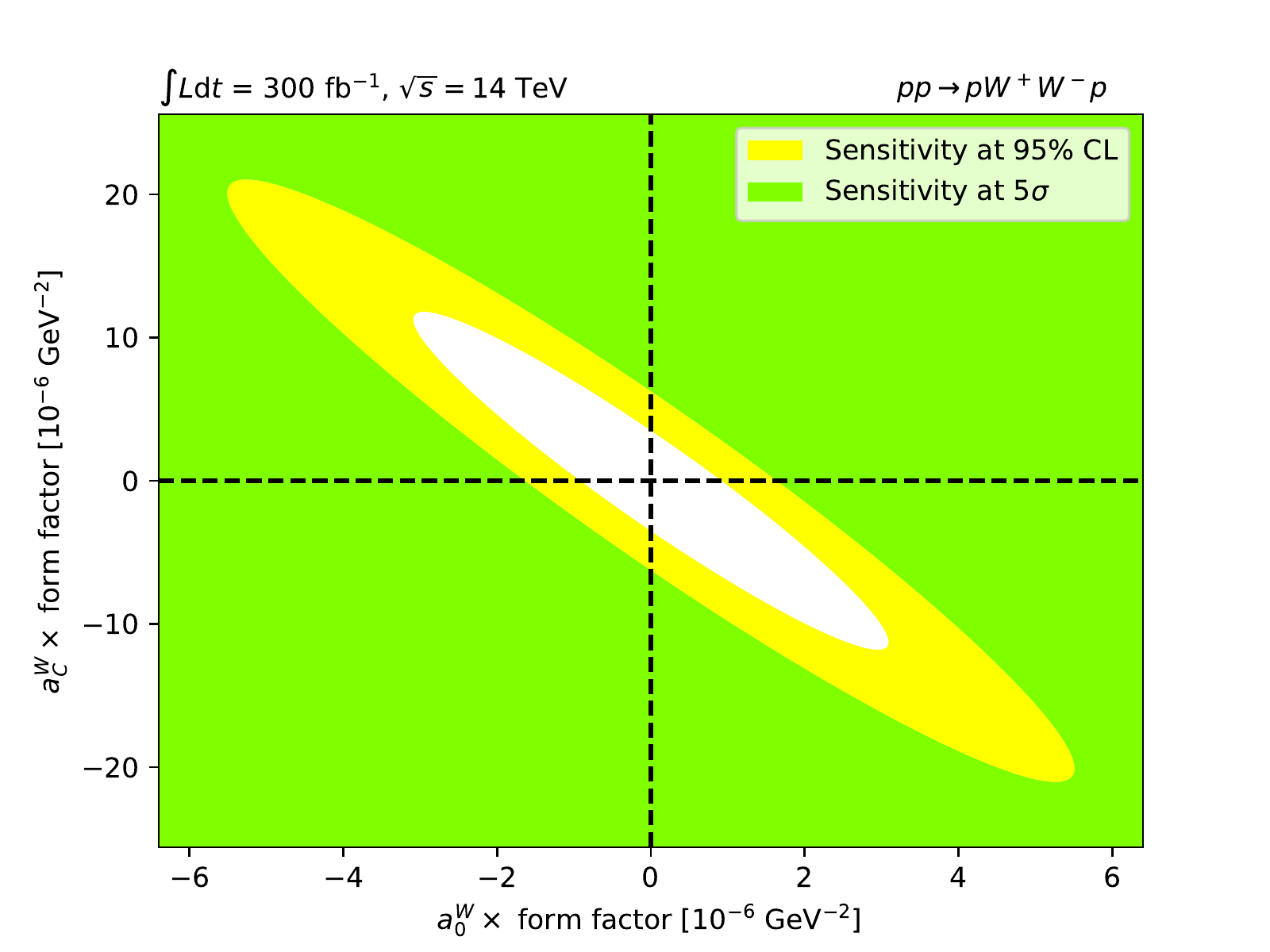}

\caption{\label{fig:a0_limit_ff} Projected sensitivities on the anomalous coupling parameters $a_0^W$ and $a_C^W$ with a dipole form factor with cutoff energy of $\Lambda_\text{cutoff} = 2$ TeV. The projections are shown for $pp\rightarrow p W^+W^- p$ at 14 TeV assuming an integrated luminosity of 300 fb$^{-1}$. The yellow and green areas represent the projected sensitivities at 95\% CL and $5\sigma$, respectively. The results correspond to a combination of hadronic, semi-leptonic, and leptonic decay channels of the $W^+W^-$ system. The blank area in the center represents the region where we do not expect sensitivity to the anomalous coupling parameter. Time-of-flight measurements with 20 ps precision is assumed.}
\end{figure}}

The ATLAS and CMS experiments set one-dimensional bounds on the interaction couplings without form factors of $|a_0^W| < 1.1 \times 10^{-6}$ GeV$^{-2}$ and $|a_C^W| < 4.1 \times 10^{-5}$ GeV$^{-2}$ at 95\% CL based on the 7 and 8 TeV analysis results~\cite{CMS_yyWW7TeV, CMS_yyWW8TeV, ATLAS_yyWW8TeV}. However, a direct comparison of our projections to existing limits published by the ATLAS and CMS experiments is not very straightforward. The reason is that, in these studies, the anomalous production of $W^+W^-$ pairs is allowed to contribute at arbitrarily large invariant masses of the diboson system. As mentioned in Sec.~\ref{sec:anomalous_coupling}, the production cross section of the $\gamma\gamma\rightarrow W^+W^-$ interaction induced by the dimension-six effective operator increases with $m_{WW}$, eventually leading to a violation of unitarity. Thus, the sensitivity is artificially enhanced in studies where no upper cut in the invariant mass of the $W$ boson pair is imposed. In our case, such an upperbound in mass is imposed due to the acceptance in $\xi$ of the protons reconstructed at the RPs. This has been noted in other studies with anomalous coupling interactions using the proton tagging technique, such as in Refs.~\cite{yyyZ_Baldenegro, yyyy_Fichet}. To avoid quoting limits on the kinematic region where unitarity violation takes place, the ATLAS and CMS Collaborations presented exclusion limits on the anomalous couplings using a dipole form factor with an energy cutoff at $\Lambda_\text{cutoff} = 500$ GeV~\cite{CMS_yyWW7TeV, CMS_yyWW8TeV, ATLAS_yyWW8TeV}. The one-dimensional limits at 95\% CL with said form factor are $|a_0^W| < 0.9\times 10^{-4}$ GeV$^{-2}$ and $|a_C^W| < 3.6 \times 10^{-4}$ GeV$^{-2}$~\cite{CMS_yyWW7TeV, CMS_yyWW8TeV, ATLAS_yyWW8TeV}. However, the energy cutoff $\Lambda_\text{cutoff} = 500$ GeV used in the CMS and ATLAS analyses is too low for the study presented in this paper, given that the minimum invariant mass of the $W$ boson pair in our study starts at 1 TeV, i.e., the cross section vanishes for the phase-space we can probe. For these reasons, a direct comparison with existing limits is not possible. No limits on anomalous couplings from the 13 TeV analysis of $\gamma^*\gamma^*\rightarrow W^+W^-$ by the ATLAS Collaboration were reported.

We draw a comparison with projections presented in the CMS-TOTEM PPS technical design report~\cite{PPS}. In these studies, only leptonic decays are considered, for $0.015 < \xi < 0.15 $. PPS expects $4\times10^{-6}$($1\times10^{-5}$) GeV$^{-2}$ for $|a_{0}^W|$ ($|a_C^W|$) in the leptonic channel alone for an integrated luminosity of 100 fb$^{-1}$ at 13 TeV~\cite{PPS}, within the same order of magnitude as those obtained here for the leptonic channel as expected. Thus, when considering the hadronic and semi-leptonic final states, we can improve this bound by an order of magnitude over those original projections. In a recent study~\cite{Tizchang:2020tqs}, it was proposed to use central exclusive $W^+W^-\gamma$ production to probe non-abelian couplings between photons and $W$ bosons of the SM. The latter can be used to constrain anomalous gauge quartic couplings, and it was found that for the dimension-six operators couplings they have a sensitivity of $|a_0^W| < 10^{-6}$ GeV$^{-2}$ at 95\% CL, i.e., similar to the sensitivity we have in leptonic decays of exclusive $W^+W^-$ events. It would be interesting to see how much the projections presented in Ref.~\cite{Tizchang:2020tqs} could be improved if semi-leptonic or hadronic decays of $W$ bosons are considered. In Ref.~\cite{Gurkanli:2020tee}, the sensitivity to the same anomalous quartic vertex is estimated for the Future Circular Collider in electron-proton mode (FCC-he) at $\sqrt{s} = 5.29$ TeV at an integrated luminosity of 1 ab$^{-1}$ of luminosity, with a focus on the process $e^- p \rightarrow e^- W^+ \gamma q X$. The sensitivity to interaction couplings $|f_{M,0}/\Lambda^4|$ associated to dimension-eight $\gamma\gamma W^+W^-$ operators is expected to be of the order of $9$ TeV$^{-4}$ at 95\% CL under the aforementioned assumptions~\cite{Gurkanli:2020tee}. This translates to a projection on $|a_0^W| < 2.3 \times 10^{-7}$ GeV$^{-2}$ at 95\% CL, with the relations between dimension-six and dimension-eight operators described in Refs.~\cite{dimension8-1,dimension8-2}. Our projections are thus competitive with these potential future collider scenarios, while ours are readily accessible at current LHC collider energies and running conditions.

A simple extrapolation (assuming similar signal-to-background ratio) of our results to 3000 fb$^{-1}$ for the High Luminosity LHC (HL-LHC) suggests an improvement of our projections by at most a factor of about $\approx$ 3 on $a_0^W$, $a_C^W$. For extracting robust projections for the HL-LHC, one would need to have a good understanding of the effects of a larger pileup environment on the reconstruction of the $W^+W^-$ boson pair decay, and of protons detected in the RP detectors. Precise time-of-flight measurements of the scattered protons will be of essence for these studies at the HL-LHC.

\begin{table}[]
\renewcommand{\arraystretch}{1.2}
\resizebox{\textwidth}{!}{%
\begin{tabular}{c||c|c|c|c|c|c|c|c}
\hline
$W^+W^-$ decay channel    & \multicolumn{2}{c|}{Hadronic} & \multicolumn{2}{c|}{Semi-leptonic} & \multicolumn{2}{c|}{Leptonic} & \multicolumn{2}{c}{Combined} \\ \hline
Coupling {[} 10$^{-7}$ GeV$^{-2}${]} & 5$\sigma$     & $95\%$ CL    & 5$\sigma$       & $95\%$ CL       & 5$\sigma$     & $95\%$ CL    & 5$\sigma$     & $95\%$ CL    \\ \hline\hline
$|a_0^W|$, $a_C^W = 0$ (no form factor)  &        6.9        &    3.8          &         10         &          4.9       &     43           &   24           &       6.6         &           3.7   \\ 
$|a_0^W|$, $a_C^W = 0 $ (form factor)     &       17         &      9.4        &         27         &       13           &     43           &   24           &         16       &     9.2         \\ \hline
$|a_C^W|$, $a_0^W = 0 $ (no form factor)  &       17         &       9.5       &           25       &           12      &         107       &   59           &         16       &        9      \\ 
$|a_C^W|$, $a_0^W = 0 $ (form factor)     &       42.0         &      23        &          67          &       33          &           107     &    59          &        41        &        23      \\ \hline
\end{tabular}
}
\caption{Sensitivities to the anomalous couplings $a_0^W$ and $a_C^W$ assuming an integrated luminosity of 300 fb$^{-1}$ and a center of mass energy of $\sqrt{s} = 14$ TeV. The sensitivities for each decay channel of the $W^+W^-$ is displayed in the columns. The combined sensitivities are presented in the last column. Sensitivities assuming no form factor and a form factor with cutoff scale $\Lambda_\text{cutoff} = 2$ TeV are displayed in the rows. These sensitivities establish the coupling values probed at 5$\sigma$ or those which could be excluded at 95\% CL.\label{final}}
\end{table}

\subsection{$|t|$ distribution in the SM and BSM}

\begin{figure}
\centering
\includegraphics[width=0.7\linewidth]{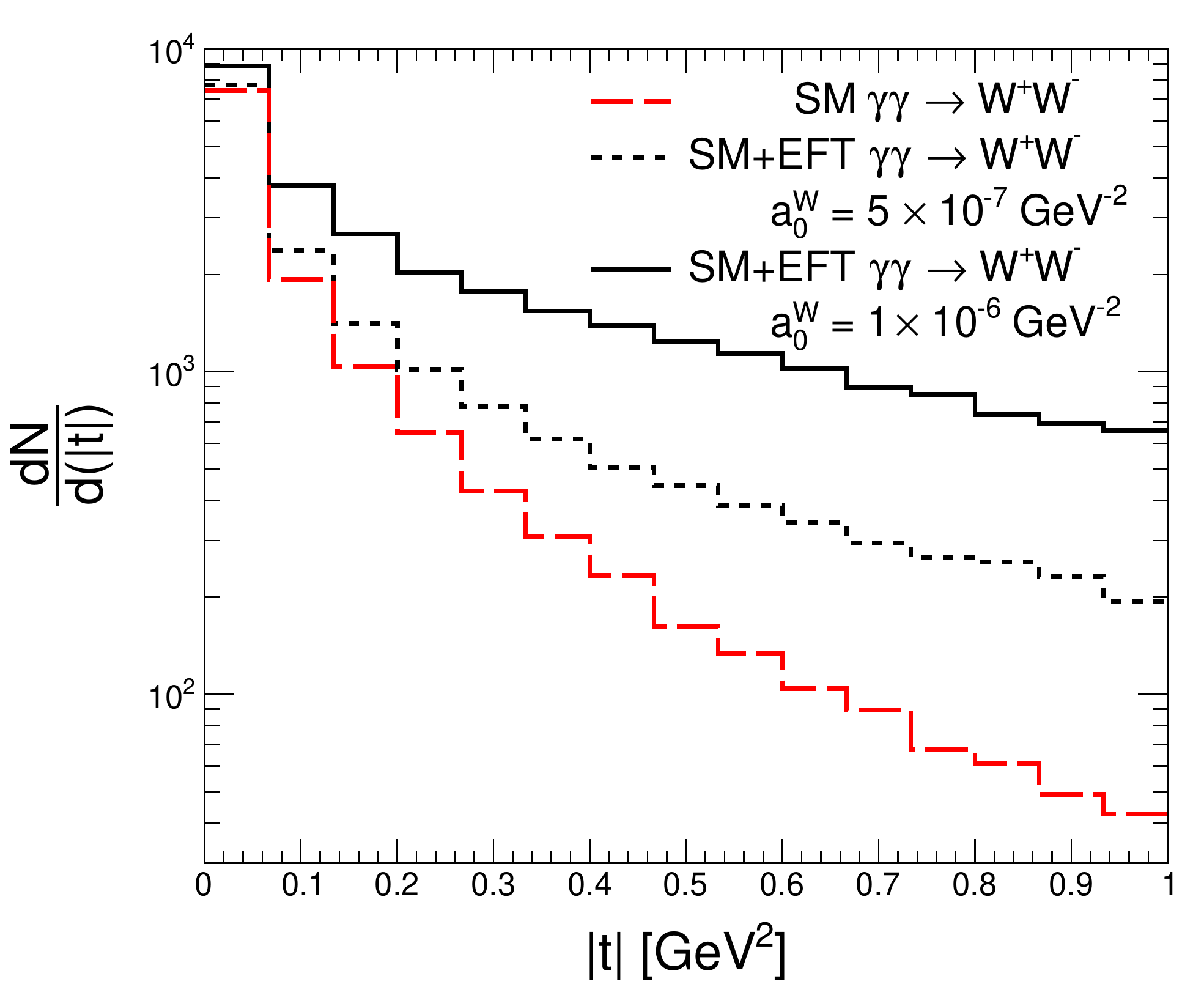}

\caption{\label{fig:t_distribution} Square of four-momentum transfer at the proton vertex, $|t|$, in $pp\rightarrow p W^+W^- p$ events at generator-level, for protons scattering towards positive or negative rapidities. The SM contribution is represented by the red histogram, while the SM and anomalous couplings distributions combined for coupling values of $a_0^W = 5 \times 10^{-7}$ GeV$^{-2}$ and $a_0^W = 1 \times 10^{-6}$ GeV$^{-2}$ are represented by the solid and dashed black line, respectively. The distributions are normalized to cross section and luminosity.}
\end{figure}

As an additional remark, we analyze the four-momentum transfer squared at the proton vertex, $ t \equiv (p_f-p_i)^2$, where $p_i$ and $p_f$ are the four-momenta of the scattered proton before and after the interaction, respectively. In Fig.~\ref{fig:t_distribution}, we show the $|t|$ distribution for $\gamma\gamma\rightarrow W^+W^-$ in the SM (red dashed line) and SM and anomalous coupling (black lines) at generator-level. Exclusive $W^+W^-$ with a contribution of anomalous couplings would yield an overall flatter $|t|$ distribution, while the SM-only prediction yields $|t| \ll 1$ GeV$^{2}$. This is due to the photon-flux in the equivalent photon approximation, where high $\xi$ is correlated with larger photon virtualities $-q^2$, which in turn directly affect $|t|$. In other words, if events at high $m_{WW}$ and high $p_T$ with two protons were observed experimentally, one could cross check the nature of said process by inspecting the values of the reconstructed $|t|$. If the protons are coming from pileup interactions, they will also have small values of $|t|$ (like the signal). Thus, it is of high importance to remove in an independent way the pileup events by matching the mass and the rapidity of the $W^+W^-$ candidate and the $pp$ system, as discussed throughout this paper. It is worth noting that the survival probability might also be $t$-dependent, so that further studies are needed to investigate the effect on SM and anomalous coupling $|t|$-distributions. The resolution on the reconstructed $|t|$ is not very good because of the larger beam divergence in standard LHC runs~\cite{PPS, AFP}. However, even if these smearing effects are large, one could use this feature as an additional check.

\section{Summary}\label{sec:summary}

Sensitivities on central exclusive production of high-mass $W^+W^-$ boson pairs in pp collisions have been studied assuming an integrated luminosity of 300 fb$^{-1}$ at $\sqrt{s} = 14 $ TeV. We based our projections on the process $pp\rightarrow p W^+W^- p$, where both scattered protons are detected with dedicated forward proton detectors installed around the interaction points of the ATLAS and CMS detectors, known as Roman Pots (RP) detectors.  This process allows us to study in detail the $\gamma\gamma \rightarrow W^+W^-$ process. In previous phenomenology and experimental studies, only the leptonic final state of the diboson system has been considered due to its clean experimental signature. In order to probe larger $W^+W^-$ invariant masses, we have to consider additional decay channels other than the standard leptonic decay. Using the forward proton detectors, one can consider every combination of final-state decays of the $W$ bosons.

We studied exclusive $W^+W^-$ production in hadronic ($W^+W^-\rightarrow \text{hadrons}$, clustered into anti-$k_t$ $ R =0.8$ jets for merged or partially merged $W\rightarrow q\bar{q}'$ decay topologies) and semi-leptonic ($W^+W^-\rightarrow \text{hadrons}+\ell\nu_\ell$, with the hadrons clustered into anti-$k_t$ $R = 0.8$ jets) final states, and compared their respective sensitivity to the standard leptonic decays of both $W$ bosons ($W^+W^-\rightarrow \ell_1 \nu_{\ell_1} \ell_2 \nu_{\ell_2}$) with two tagged protons. In the hadronic channel, a measurement of the time-of-flight difference of the scattered protons, together with a judicious choice jet substructure variables, will be instrumental to further control the multijet background created in quark and gluon standard quantum chromodynamics interactions. The semi-leptonic channel offers a good compromise in terms of signal-to-background ratio with rather basic event selection requirements. Here, the dominant background comes from non-diffractive $W^\pm$+jet background in association with pileup protons. The leptonic channel provides access mostly to lower masses $m_{WW}$ and softer $p_T^W$, and is thus complementary to the kinematic reach of the hadronic and semi-leptonic channels. The various $W^+W^-$ decay channels, being mutually exclusive, can be used independently as standard candle processes to further cross check the RP detector calibration, in addition to $\gamma\gamma\rightarrow \ell^+\ell^-$ scattering already used by the ATLAS and CMS Collaborations.

In addition to the aforementioned study, we considered scenarios where deviations from the Standard Model expectations are present. We parametrize these deviations by means of an anomalous $\gamma\gamma WW$ vertex in a dimension-six effective field theory framework. Here, the hadronic final state gives much better sensitivity to beyond SM deviations at high-mass and high-$p_T$ of each $W$ boson. The hadronic channel yields the best sensitivity, followed by the semi-leptonic and leptonic channels. This is because $\gamma\gamma\rightarrow W^+W^-$ with anomalous coupling contributions are more likely to yield the boosted topology necessary to establish large-radius jet reconstruction, and because the QCD jet background is further suppressed at large $m_{WW}$ and larger $p_T^W$. The interaction couplings $a_0^W$ and $a_C^W$ can be probed down to values of $3.7 \times 10^{-7}$ GeV$^{-2}$ and $9.2\times 10^{-7}$ GeV$^{-2}$ at 95\% CL by combining all decay channels (one-dimensional projections, while keeping one of the coupling parameters fixed to zero, without form factors). This is contrast to the sensitivity reached with the leptonic channel alone of $4.3 \times 10^{-6}$ GeV$^{-2}$ and $10^{-5}$ GeV$^{-2}$ at 95\% CL. Comparisons to sensitivities to the same quartic interaction coupling probed in different processes are made.

\acknowledgments

We thank Maxime Massa who helped us in the initial phase of the anomalous coupling study. We also thank Sylvain Fichet for his helpful feedback on the manuscript. This work has been financially supported by the starting grant of the Foundation Distinguished Professor program at the University of Kansas.

\bibliographystyle{JHEP} 
\bibliography{references.bib}

\providecommand{\href}[2]{#2}\begingroup\raggedright\begin{thebibliography}{10}

\bibitem{AQGC_FichetGersdorff}
S.~Fichet and G.~von Gersdorff, {\it {Anomalous gauge couplings from composite
  Higgs and warped extra dimensions}},  {\em JHEP} {\bf 03} (2014) 102,
  [\href{http://arxiv.org/abs/1311.6815}{{\tt arXiv:1311.6815}}].

\bibitem{EFT1}
B.~Grzadkowski, M.~Iskrzynski, M.~Misiak, and J.~Rosiek, {\it {Dimension-Six
  Terms in the Standard Model Lagrangian}},  {\em JHEP} {\bf 10} (2010) 085,
  [\href{http://arxiv.org/abs/1008.4884}{{\tt arXiv:1008.4884}}].

\bibitem{Espriu:2014jya}
D.~Espriu and F.~Mescia, {\it {Unitarity and causality constraints in composite
  Higgs models}},  {\em Phys. Rev. D} {\bf 90} (2014), no.~1 015035,
  [\href{http://arxiv.org/abs/1403.7386}{{\tt arXiv:1403.7386}}].

\bibitem{Delgado:2014jda}
R.~Delgado, A.~Dobado, M.~Herrero, and J.~Sanz-Cillero, {\it {One-loop
  $\gamma\gamma \to$ W$_{L}^{+}$ W$_{L}^{-}$ and $\gamma\gamma \to$ Z$_{L}$
  Z$_{L}$ from the Electroweak Chiral Lagrangian with a light Higgs-like
  scalar}},  {\em JHEP} {\bf 07} (2014) 149,
  [\href{http://arxiv.org/abs/1404.2866}{{\tt arXiv:1404.2866}}].

\bibitem{VBSreview}
A.~B. et~al., {\it {Vector boson scattering: Recent experimental and theory
  developments}},  {\em Reviews in Physics} {\bf 3} (2018) 44.

\bibitem{Gallinaro:2020cte}
M.~Gallinaro et~al., {\it {Beyond the Standard Model in Vector Boson Scattering
  Signatures}},  2020.
\newblock \href{http://arxiv.org/abs/2005.09889}{{\tt arXiv:2005.09889}}.

\bibitem{WeiszackerWilliams}
C.~Carimalo, P.~Kessler, and J.~Parisi, {\it {Validity of the equivalent-photon
  approximation for virtual photon-photon collisions}},  {\em Phys. Rev. D}
  {\bf 20} (1979) 1057--1068.

\bibitem{Budnev}
V.~Budnev, I.~Ginzburg, G.~Meledin, and V.~Serbo, {\it {The Two photon particle
  production mechanism. Physical problems. Applications. Equivalent photon
  approximation}},  {\em Phys. Rept.} {\bf 15} (1975) 181--281.

\bibitem{AFP}
L.~Adamczyk, E.~Banaś, A.~Brandt, M.~Bruschi, S.~Grinstein, J.~Lange,
  M.~Rijssenbeek, P.~Sicho, R.~Staszewski, T.~Sykora, M.~Trzebiński,
  J.~Chwastowski, and K.~Korcyl, {\it {Technical Design Report for the ATLAS
  Forward Proton Detector}},  Tech. Rep. CERN-LHCC-2015-009. ATLAS-TDR-024,
  2015.

\bibitem{PPS}
M.~Albrow, M.~Arneodo, V.~Avati, J.~Baechler, N.~Cartiglia, M.~Deile,
  M.~Gallinaro, J.~Hollar, M.~Lo~Vetere, K.~Oesterberg, N.~Turini, J.~Varela,
  D.~Wright, and C.~CMS-TOTEM, {\it {CMS-TOTEM Precision Proton Spectrometer}},
   Tech. Rep. CERN-LHCC-2014-021. TOTEM-TDR-003. CMS-TDR-13, 2014.

\bibitem{ggWW_ChaponKepkaRoyon}
E.~Chapon, C.~Royon, and O.~Kepka, {\it {Anomalous quartic $W W \gamma \gamma$,
  $Z Z \gamma \gamma$, and trilinear $WW \gamma$ couplings in two-photon
  processes at high luminosity at the LHC}},  {\em Phys. Rev. D} {\bf 81}
  (2010) 074003, [\href{http://arxiv.org/abs/0912.5161}{{\tt
  arXiv:0912.5161}}].

\bibitem{PhysRevD.78.073005}
O.~Kepka and C.~Royon, {\it {Anomalous $WW\gamma$ coupling in photon-induced
  processes using forward detectors at the CERN LHC}},  {\em Phys. Rev. D} {\bf
  78} (2008) 073005.

\bibitem{yyyZ_Baldenegro}
C.~Baldenegro, S.~Fichet, G.~von Gersdorff, and C.~Royon, {\it {Probing the
  anomalous $\gamma$$\gamma$$\gamma$Z coupling at the LHC with proton
  tagging}},  {\em JHEP} {\bf 06} (2017) 142,
  [\href{http://arxiv.org/abs/1703.10600}{{\tt arXiv:1703.10600}}].

\bibitem{yyyy_Fichet}
S.~Fichet, G.~von Gersdorff, B.~Lenzi, C.~Royon, and M.~Saimpert, {\it
  {Light-by-light scattering with intact protons at the LHC: from Standard
  Model to New Physics}},  {\em JHEP} {\bf 02} (2015) 165,
  [\href{http://arxiv.org/abs/1411.6629}{{\tt arXiv:1411.6629}}].

\bibitem{Fichet:2016clq}
S.~Fichet, {\it {Shining Light on Polarizable Dark Particles}},  {\em JHEP}
  {\bf 04} (2017) 088, [\href{http://arxiv.org/abs/1609.01762}{{\tt
  arXiv:1609.01762}}].

\bibitem{Inan:2014mua}
S.~\.Inan, {\it {Dimension-six anomalous $tq\gamma$ couplings in $\gamma\gamma$
  collision at the LHC}},  {\em Nucl. Phys. B} {\bf 897} (2015) 289--301,
  [\href{http://arxiv.org/abs/1410.3609}{{\tt arXiv:1410.3609}}].

\bibitem{Senol:2014vta}
A.~Senol and M.~Köksal, {\it {Analysis of anomalous quartic $WWZ\gamma$
  couplings in $\gamma p$ collision at the LHC}},  {\em Phys. Lett. B} {\bf
  742} (2015) 143--148, [\href{http://arxiv.org/abs/1410.3648}{{\tt
  arXiv:1410.3648}}].

\bibitem{Fichet:2016pvq}
S.~Fichet, G.~von Gersdorff, and C.~Royon, {\it {Measuring the Diphoton
  Coupling of a 750 GeV Resonance}},  {\em Phys. Rev. Lett.} {\bf 116} (2016),
  no.~23 231801, [\href{http://arxiv.org/abs/1601.01712}{{\tt
  arXiv:1601.01712}}].

\bibitem{Gupta:2011be}
R.~S. Gupta, {\it {Probing Quartic Neutral Gauge Boson Couplings using
  diffractive photon fusion at the LHC}},  {\em Phys. Rev. D} {\bf 85} (2012)
  014006, [\href{http://arxiv.org/abs/1111.3354}{{\tt arXiv:1111.3354}}].

\bibitem{Senol:2013ym}
A.~Senol, {\it {$ZZ\gamma$ and $Z\gamma\gamma$ anomalous couplings in $\gamma
  p$ collision at the LHC}},  {\em Phys. Rev. D} {\bf 87} (2013) 073003,
  [\href{http://arxiv.org/abs/1301.6914}{{\tt arXiv:1301.6914}}].

\bibitem{Sun:2014qoa}
H.~Sun, {\it {Probe anomalous tq$\gamma$ couplings through single top
  photoproduction at the LHC}},  {\em Nucl. Phys. B} {\bf 886} (2014) 691--711,
  [\href{http://arxiv.org/abs/1402.1817}{{\tt arXiv:1402.1817}}].

\bibitem{Fichet:2013gsa}
S.~Fichet, G.~von Gersdorff, O.~Kepka, B.~Lenzi, C.~Royon, and M.~Saimpert,
  {\it {Probing new physics in diphoton production with proton tagging at the
  Large Hadron Collider}},  {\em Phys. Rev. D} {\bf 89} (2014) 114004,
  [\href{http://arxiv.org/abs/1312.5153}{{\tt arXiv:1312.5153}}].

\bibitem{Sun:2014qba}
H.~Sun, {\it {Large Extra Dimension effects through Light-by-Light Scattering
  at the CERN LHC}},  {\em Eur. Phys. J. C} {\bf 74} (2014), no.~8 2977,
  [\href{http://arxiv.org/abs/1406.3897}{{\tt arXiv:1406.3897}}].

\bibitem{TAHERIMONFARED2016301}
S.~Taheri-Monfared, S.~Fayazbakhsh, and M.~Mohammadi-Najafabadi, {\it
  {Exploring anomalous $HZ\gamma$ couplings in $\gamma$-proton collisions at
  the LHC}},  {\em Physics Letters B} {\bf 762} (2016) 301 -- 308.

\bibitem{Gurkanli:2020tee}
E.~Gurkanli, V.~Ari, A.~Gutiérrez-Rodríguez, M.~Hernández-Ruíz, and
  M.~Koksal, {\it {Sensitivity physics expected to the measurement of the
  quartic $WW\gamma\gamma$ couplings at the LHeC and the FCC-he}},
  \href{http://arxiv.org/abs/2003.06669}{{\tt arXiv:2003.06669}}.

\bibitem{Sylvain_PRD}
S.~Fichet, G.~von Gersdorff, O.~Kepka, B.~Lenzi, C.~Royon, and M.~Saimpert,
  {\it {Probing new physics in diphoton production with proton tagging at the
  Large Hadron Collider}},  {\em Phys. Rev. D} {\bf 89} (2014) 114004.

\bibitem{Baldenegro_ALP}
C.~Baldenegro, S.~Fichet, G.~von Gersdorff, and C.~Royon, {\it {Searching for
  axion-like particles with proton tagging at the LHC}},  {\em JHEP} {\bf 06}
  (2018) 131, [\href{http://arxiv.org/abs/1803.10835}{{\tt arXiv:1803.10835}}].

\bibitem{Goncalves:2020saa}
V.~P. Gon\c{c}alves, D.~E. Martins, M.~S. Rangel, and M.~Tasevsky, {\it {Top
  quark pair production in the exclusive processes at LHC}},
  \href{http://arxiv.org/abs/2007.04565}{{\tt arXiv:2007.04565}}.

\bibitem{Harland-Lang:2020veo}
L.~Harland-Lang, M.~Tasevsky, V.~Khoze, and M.~Ryskin, {\it {A new approach to
  modelling elastic and inelastic photon-initiated production at the LHC:
  SuperChic 4}},  \href{http://arxiv.org/abs/2007.12704}{{\tt
  arXiv:2007.12704}}.

\bibitem{Szczurek:2019ihz}
A.~Szczurek and M.~{\L}uszczak, {\it {Production of $W^+ W^-$ and $t \bar t$
  pairs via photon-photon processes in proton-proton scattering}},  {\em PoS}
  {\bf DIS2019} (2019) 035, [\href{http://arxiv.org/abs/1907.08936}{{\tt
  arXiv:1907.08936}}].

\bibitem{Luszczak:2018ntp}
M.~{\L}uszczak, W.~Sch\"afer, and A.~Szczurek, {\it {Production of $W^+ W^-$
  pairs via $\gamma^*\gamma^* \to W^+ W^-$ subprocess with photon transverse
  momenta}},  {\em JHEP} {\bf 05} (2018) 064,
  [\href{http://arxiv.org/abs/1802.03244}{{\tt arXiv:1802.03244}}].

\bibitem{Forthomme:2018sxa}
L.~Forthomme, M.~{\L}uszczak, W.~Sch\"afer, and A.~Szczurek, {\it {Rapidity gap
  survival factors caused by remnant fragmentation for $W^+ W^-$ pair
  production via $\gamma^*\gamma^* \to W^+ W^-$ subprocess with photon
  transverse momenta}},  {\em Phys. Lett. B} {\bf 789} (2019) 300--307,
  [\href{http://arxiv.org/abs/1805.07124}{{\tt arXiv:1805.07124}}].

\bibitem{dilepton_PPS}
{\bf CMS, TOTEM} Collaboration, A.~M. Sirunyan et~al., {\it {Observation of
  proton-tagged, central (semi)exclusive production of high-mass lepton pairs
  in pp collisions at 13 TeV with the CMS-TOTEM precision proton
  spectrometer}},  {\em JHEP} {\bf 07} (2018) 153,
  [\href{http://arxiv.org/abs/1803.04496}{{\tt arXiv:1803.04496}}].

\bibitem{dilepton_AFP}
{\bf ATLAS} Collaboration, {\it {Observation and measurement of forward proton
  scattering in association with lepton pairs produced via the photon fusion
  mechanism at ATLAS}}, .

\bibitem{exclusive_diphoton_PPS}
CMS and TOTEM, {\it {First search for exclusive diphoton production at high
  mass with intact protons in proton-proton collisions at √ s =13 TeV at the
  LHC}},  Tech. Rep. CMS-PAS-EXO-18-014, 2020.

\bibitem{WWyy_LEP}
{\bf OPAL} Collaboration, G.~Abbiendi et~al., {\it {A study of
  $W^{+}W^{-}\gamma$ events at LEP}},  {\em Phys. Lett. B} {\bf 580} (2004) 17,
  [\href{http://arxiv.org/abs/hep-ex/0309013}{{\tt hep-ex/0309013}}].

\bibitem{WWyy_LEP2}
{\bf OPAL} Collaboration, G.~Abbiendi et~al., {\it {Measurement of the
  $W^{+}W^{-}\gamma$ cross-section and first direct limits on anomalous
  electroweak quartic gauge couplings}},  {\em Phys. Lett. B} {\bf 471} (1999)
  293, [\href{http://arxiv.org/abs/hep-ex/9910069}{{\tt hep-ex/9910069}}].

\bibitem{WWyy_LEP3}
{\bf DELPHI} Collaboration, J.~Abdallah et~al., {\it {Measurement of the
  e$^{+}$e$^{-}\to W^{+}W^{-}\gamma$ cross-section and limits on anomalous
  quartic gauge couplings with {DELPHI}}},  {\em Eur. Phys. J. C} {\bf 31}
  (2003) 139, [\href{http://arxiv.org/abs/hep-ex/0311004}{{\tt
  hep-ex/0311004}}].

\bibitem{WWyy_LEP4}
{\bf L3} Collaboration, P.~Achard et~al., {\it {Study of the $W^{+}W^{-}\gamma$
  process and limits on anomalous quartic gauge boson couplings at {LEP}}},
  {\em Phys. Lett. B} {\bf 527} (2002) 29,
  [\href{http://arxiv.org/abs/hep-ex/0111029}{{\tt hep-ex/0111029}}].

\bibitem{Abazov:2013opa}
{\bf D0} Collaboration, V.~M. Abazov et~al., {\it {Search for anomalous quartic
  $WW{\gamma}{\gamma}$ couplings in dielectron and missing energy final states
  in $p\bar{p}$ collisions at $\sqrt{s}$ = 1.96 TeV}},  {\em Phys. Rev. D} {\bf
  88} (2013) 012005, [\href{http://arxiv.org/abs/1305.1258}{{\tt
  arXiv:1305.1258}}].

\bibitem{CMS_yyWW7TeV}
{\bf CMS} Collaboration, S.~Chatrchyan et~al., {\it {Study of Exclusive
  Two-Photon Production of $W^+W^-$ in $pp$ Collisions at $\sqrt{s} = 7$ TeV
  and Constraints on Anomalous Quartic Gauge Couplings}},  {\em JHEP} {\bf 07}
  (2013) 116, [\href{http://arxiv.org/abs/1305.5596}{{\tt arXiv:1305.5596}}].

\bibitem{CMS_yyWW8TeV}
{\bf CMS} Collaboration, V.~Khachatryan et~al., {\it {Evidence for exclusive
  $\gamma\gamma \to W^+ W^-$ production and constraints on anomalous quartic
  gauge couplings in $pp$ collisions at $ \sqrt{s}=7 $ and 8 TeV}},  {\em JHEP}
  {\bf 08} (2016) 119, [\href{http://arxiv.org/abs/1604.04464}{{\tt
  arXiv:1604.04464}}].

\bibitem{ATLAS_yyWW8TeV}
{\bf ATLAS} Collaboration, M.~Aaboud et~al., {\it {Measurement of exclusive
  $\gamma\gamma\rightarrow W^+W^-$ production and search for exclusive Higgs
  boson production in $pp$ collisions at $\sqrt{s} = 8$ TeV using the ATLAS
  detector}},  {\em Phys. Rev. D} {\bf 94} (2016), no.~3 032011,
  [\href{http://arxiv.org/abs/1607.03745}{{\tt arXiv:1607.03745}}].

\bibitem{gammagammaWW_ATLASCONF}
{\bf {ATLAS Collaboration}} Collaboration, {\it {Observation of photon-induced
  $W^+W^-$ production in $pp$ collisions at $\sqrt{s}=13$ TeV using the ATLAS
  detector}},  Tech. Rep. ATLAS-CONF-2020-038, CERN, Geneva, 2020.

\bibitem{Terazawa}
H.~Terazawa, {\it {Two photon processes for particle production at
  high-energies}},  {\em Rev. Mod. Phys.} {\bf 45} (1973) 615--662.

\bibitem{Khoze1}
V.~Khoze, A.~Martin, and M.~Ryskin, {\it {Prospects for new physics
  observations in diffractive processes at the LHC and Tevatron}},  {\em Eur.
  Phys. J. C} {\bf 23} (2002) 311--327,
  [\href{http://arxiv.org/abs/hep-ph/0111078}{{\tt hep-ph/0111078}}].

\bibitem{Khoze2}
V.~Khoze, A.~Martin, and M.~Ryskin, {\it {Photon exchange processes at hadron
  colliders as a probe of the dynamics of diffraction}},  {\em Eur. Phys. J. C}
  {\bf 24} (2002) 459--468, [\href{http://arxiv.org/abs/hep-ph/0201301}{{\tt
  hep-ph/0201301}}].

\bibitem{survival_khoze}
V.~Khoze, A.~Martin, and M.~Ryskin, {\it {Multiple interactions and rapidity
  gap survival}},  {\em J. Phys. G} {\bf 45} (2018), no.~5 053002,
  [\href{http://arxiv.org/abs/1710.11505}{{\tt arXiv:1710.11505}}].

\bibitem{survival_superchic2}
L.~Harland-Lang, V.~Khoze, and M.~Ryskin, {\it {Exclusive physics at the LHC
  with SuperChic 2}},  {\em Eur. Phys. J. C} {\bf 76} (2016), no.~1 9,
  [\href{http://arxiv.org/abs/1508.02718}{{\tt arXiv:1508.02718}}].

\bibitem{survival_gotsman}
E.~Gotsman, E.~Levin, U.~Maor, E.~Naftali, and A.~Prygarin, {\it {Survival
  probability of large rapidity gaps}},  in {\em {HERA and the LHC: A Workshop
  on the Implications of HERA for LHC Physics: CERN - DESY Workshop 2004/2005
  (Midterm Meeting, CERN, 11-13 October 2004; Final Meeting, DESY, 17-21
  January 2005)}}, pp.~221--241, 2005.
\newblock \href{http://arxiv.org/abs/hep-ph/0511060}{{\tt hep-ph/0511060}}.

\bibitem{Cacciari:2008gp}
M.~Cacciari, G.~P. Salam, and G.~Soyez, {\it {The anti-$k_t$ jet clustering
  algorithm}},  {\em JHEP} {\bf 04} (2008) 063,
  [\href{http://arxiv.org/abs/0802.1189}{{\tt arXiv:0802.1189}}].

\bibitem{fastjet}
M.~Cacciari, G.~P. Salam, and G.~Soyez, {\it {FastJet User Manual}},  {\em Eur.
  Phys. J. C} {\bf 72} (2012) 1896, [\href{http://arxiv.org/abs/1111.6097}{{\tt
  arXiv:1111.6097}}].

\bibitem{PFnoteCMS}
{\bf CMS} Collaboration, A.~Sirunyan et~al., {\it {Particle-flow reconstruction
  and global event description with the CMS detector}},  {\em JINST} {\bf 12}
  (2017), no.~10 P10003, [\href{http://arxiv.org/abs/1706.04965}{{\tt
  arXiv:1706.04965}}].

\bibitem{PFnoteATLAS}
{\bf ATLAS} Collaboration, M.~Aaboud et~al., {\it {Jet reconstruction and
  performance using particle flow with the ATLAS Detector}},  {\em Eur. Phys.
  J. C} {\bf 77} (2017), no.~7 466,
  [\href{http://arxiv.org/abs/1703.10485}{{\tt arXiv:1703.10485}}].

\bibitem{WbosonBoostedATLAS}
{\bf ATLAS} Collaboration, G.~Aad et~al., {\it {Measurement of the
  cross-section of high transverse momentum vector bosons reconstructed as
  single jets and studies of jet substructure in $pp$ collisions at
  ${\sqrt{s}}$ = 7 TeV with the ATLAS detector}},  {\em New J. Phys.} {\bf 16}
  (2014), no.~11 113013, [\href{http://arxiv.org/abs/1407.0800}{{\tt
  arXiv:1407.0800}}].

\bibitem{WbosonBoostedATLAS2}
{\it {Measurement of large radius jet mass reconstruction performance at
  $\sqrt{s}=8$ TeV using the ATLAS detector}},  Tech. Rep. ATLAS-CONF-2016-008,
  CERN, Geneva, 2016.

\bibitem{WbosonBoostedCMS}
{\bf CMS} Collaboration, V.~Khachatryan et~al., {\it {Identification techniques
  for highly boosted W bosons that decay into hadrons}},  {\em JHEP} {\bf 12}
  (2014) 017, [\href{http://arxiv.org/abs/1410.4227}{{\tt arXiv:1410.4227}}].

\bibitem{leptonATLAS}
{\bf ATLAS} Collaboration, G.~Aad et~al., {\it {Muon reconstruction performance
  of the ATLAS detector in proton--proton collision data at $\sqrt{s}$ =13
  TeV}},  {\em Eur. Phys. J. C} {\bf 76} (2016), no.~5 292,
  [\href{http://arxiv.org/abs/1603.05598}{{\tt arXiv:1603.05598}}].

\bibitem{leptonCMS}
{\bf CMS} Collaboration, A.~Sirunyan et~al., {\it {Performance of the CMS muon
  detector and muon reconstruction with proton-proton collisions at $\sqrt{s}=$
  13 TeV}},  {\em JINST} {\bf 13} (2018), no.~06 P06015,
  [\href{http://arxiv.org/abs/1804.04528}{{\tt arXiv:1804.04528}}].

\bibitem{MET_ATLAS}
{\bf ATLAS} Collaboration, M.~Aaboud et~al., {\it {Performance of missing
  transverse momentum reconstruction with the ATLAS detector using
  proton-proton collisions at $\sqrt{s}$ = 13 TeV}},  {\em Eur. Phys. J. C}
  {\bf 78} (2018), no.~11 903, [\href{http://arxiv.org/abs/1802.08168}{{\tt
  arXiv:1802.08168}}].

\bibitem{jetSubstructureReview}
R.~Kogler et~al., {\it {Jet Substructure at the Large Hadron Collider:
  Experimental Review}},  {\em Rev. Mod. Phys.} {\bf 91} (2019), no.~4 045003,
  [\href{http://arxiv.org/abs/1803.06991}{{\tt arXiv:1803.06991}}].

\bibitem{fpmc}
M.~Boonekamp, A.~Dechambre, V.~Juranek, O.~Kepka, M.~Rangel, C.~Royon, and
  R.~Staszewski, {\it {FPMC: A Generator for forward physics}},
  \href{http://arxiv.org/abs/1102.2531}{{\tt arXiv:1102.2531}}.

\bibitem{calchep}
A.~Belyaev, N.~D. Christensen, and A.~Pukhov, {\it {CalcHEP 3.4 for collider
  physics within and beyond the Standard Model}},  {\em Comput. Phys. Commun.}
  {\bf 184} (2013) 1729--1769, [\href{http://arxiv.org/abs/1207.6082}{{\tt
  arXiv:1207.6082}}].

\bibitem{Kepka:2008yx}
O.~Kepka and C.~Royon, {\it {Anomalous $W W \gamma$ coupling in photon-induced
  processes using forward detectors at the LHC}},  {\em Phys. Rev. D} {\bf 78}
  (2008) 073005, [\href{http://arxiv.org/abs/0808.0322}{{\tt
  arXiv:0808.0322}}].

\bibitem{herwig6}
G.~Corcella, I.~Knowles, G.~Marchesini, S.~Moretti, K.~Odagiri, P.~Richardson,
  M.~Seymour, and B.~Webber, {\it {HERWIG 6: An Event generator for hadron
  emission reactions with interfering gluons (including supersymmetric
  processes)}},  {\em JHEP} {\bf 01} (2001) 010,
  [\href{http://arxiv.org/abs/hep-ph/0011363}{{\tt hep-ph/0011363}}].

\bibitem{pythia8}
T.~Sjostrand, S.~Mrenna, and P.~Z. Skands, {\it {A Brief Introduction to PYTHIA
  8.1}},  {\em Comput. Phys. Commun.} {\bf 178} (2008) 852,
  [\href{http://arxiv.org/abs/0710.3820}{{\tt arXiv:0710.3820}}].

\bibitem{timing_detectors_saimpert}
C.~Royon, M.~Saimpert, R.~Zlebik, and O.~Kepka, {\it {Timing detectors for
  proton tagging at the LHC}},  {\em Acta Polonica B Proceedings Supplement}
  {\bf 7} (2014) 735.

\bibitem{Thaler:2010tr}
J.~Thaler and K.~Van~Tilburg, {\it {Identifying Boosted Objects with
  N-subjettiness}},  {\em JHEP} {\bf 03} (2011) 015,
  [\href{http://arxiv.org/abs/1011.2268}{{\tt arXiv:1011.2268}}].

\bibitem{largeRjetATLAS}
{\bf ATLAS Collaboration} Collaboration, {\it {Optimisation of large-radius jet
  reconstruction for the ATLAS detector in 13 TeV proton-proton collisions}},
  Tech. Rep. ATLAS-CONF-2020-021, CERN, Geneva, 2020.

\bibitem{CDF_neutrino}
{\bf CDF} Collaboration, T.~Aaltonen et~al., {\it {Direct Measurement of the
  $W$ Production Charge Asymmetry in $p\bar{p}$ Collisions at $\sqrt{s} = 1.96$
  TeV}},  {\em Phys. Rev. Lett.} {\bf 102} (2009) 181801,
  [\href{http://arxiv.org/abs/0901.2169}{{\tt arXiv:0901.2169}}].

\bibitem{EFT2}
O.~J. Eboli, M.~Gonzalez-Garcia, S.~Lietti, and S.~Novaes, {\it {Anomalous
  quartic gauge boson couplings at hadron colliders}},  {\em Phys. Rev. D} {\bf
  63} (2001) 075008, [\href{http://arxiv.org/abs/hep-ph/0009262}{{\tt
  hep-ph/0009262}}].

\bibitem{dimension8-1}
M.~Baak et~al., {\it {Working Group Report: Precision Study of Electroweak
  Interactions}},  in {\em {Community Summer Study 2013}: {Snowmass on the
  Mississippi}}, 2013.
\newblock \href{http://arxiv.org/abs/1310.6708}{{\tt arXiv:1310.6708}}.

\bibitem{dimension8-2}
G.~Belanger, F.~Boudjema, Y.~Kurihara, D.~Perret-Gallix, and A.~Semenov, {\it
  {Bosonic quartic couplings at LEP-2}},  {\em Eur. Phys. J. C} {\bf 13} (2000)
  283--293, [\href{http://arxiv.org/abs/hep-ph/9908254}{{\tt hep-ph/9908254}}].

\bibitem{statistics}
O.~Behnke, K.~Kröninger, T.~Schörner-Sadenius, and G.~Schott, eds., {\em
  {Data analysis in high energy physics}: {A practical guide to statistical
  methods}}.
\newblock Wiley-VCH, Weinheim, Germany, 2013.

\bibitem{Tizchang:2020tqs}
S.~Tizchang and S.~M. Etesami, {\it {Pinning down the gauge boson couplings in
  $WW\gamma$ production using forward proton tagging}},
  \href{http://arxiv.org/abs/2004.12203}{{\tt arXiv:2004.12203}}.

\end{thebibliography}\endgroup

\printindex
\end{document}